\documentclass[11pt]{article}

\usepackage[preprint]{acl}

\usepackage{times}
\usepackage{ulem}
\usepackage{latexsym}
\usepackage{booktabs}
\usepackage{multirow}
\usepackage{tcolorbox}
\usepackage{subcaption}
\tcbuselibrary{theorems}
\usepackage{setspace}

\usepackage[T1]{fontenc}

\usepackage[utf8]{inputenc}

\usepackage{microtype}

\usepackage{inconsolata}

\usepackage{graphicx}

\title{
    \begin{center}
        \raisebox{-0.8ex}{\includegraphics[height=1.5em]{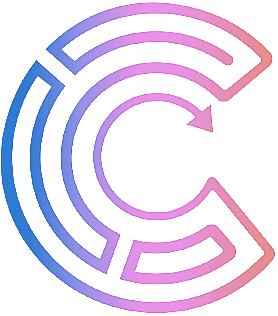}}ipherBank: Exploring the Boundary of LLM Reasoning Capabilities through Cryptography Challenges
    \end{center}
}

\author{
    Yu Li\textsuperscript{1,2},
    Qizhi Pei\textsuperscript{1,3},
    Mengyuan Sun\textsuperscript{1,2},
    Honglin Lin\textsuperscript{1,4}, \\
    {\bf Chenlin Ming\textsuperscript{1,5}},
    {\bf Xin Gao\textsuperscript{1}},
    {\bf Jiang Wu\textsuperscript{1}},
    {\bf Conghui He\textsuperscript{1}},
    {\bf Lijun Wu\textsuperscript{1}}\thanks{\ \ Corresponding author} \\
    \textsuperscript{1}Shanghai AI Laboratory \quad \textsuperscript{2}Wuhan University \quad
    \textsuperscript{3}Renmin University of China \\
    \textsuperscript{4}Beijing University of Posts and Telecommunications \quad
    \textsuperscript{5}Shanghai Jiao Tong University \\
    \texttt{\{liyu1,heconghui,wulijun\}@pjlab.org.cn} \\
    Project Page: \href{https://cipherbankeva.github.io}{https://cipherbankeva.github.io}
}

\definecolor{bluex}{rgb}{0.27, 0.42, 0.81}
\definecolor{green1}{HTML}{f3eedd}
\definecolor{purple1}{HTML}{303030}

\newtcbtheorem[number within=section]{exmp}{Example}%
{beforeafter skip balanced=.01in,colback=green1!30,colframe=purple1!70,fonttitle=\small \bfseries, left=.03in, right=.03in,bottom=.02in, top=.02in}{exmp}

\newtcbtheorem[number within=section]{exmp2}{Example}%
{beforeafter skip balanced=.01in,colback=white!50,colframe=bluex!60,fonttitle=\small \bfseries, left=.03in, right=.03in,bottom=.02in, top=.02in}{exmp2}

\begin{document}
\maketitle
\begin{abstract}

Large language models (LLMs) have demonstrated remarkable capabilities, especially the recent advancements in reasoning, such as o1 and o3, pushing the boundaries of AI. Despite these impressive achievements in mathematics and coding, the reasoning abilities of LLMs in domains requiring cryptographic expertise remain underexplored.
In this paper, we introduce \textbf{CipherBank}, a comprehensive benchmark designed to evaluate the reasoning capabilities of LLMs in cryptographic decryption tasks. 
CipherBank comprises \textbf{2,358} meticulously crafted problems, covering \textbf{262 unique plaintexts} across \textbf{5 domains} and \textbf{14 subdomains}, with a focus on privacy-sensitive and real-world scenarios that necessitate encryption. From a cryptographic perspective, CipherBank incorporates \textbf{3 major categories} of encryption methods, spanning \textbf{9 distinct algorithms}, ranging from classical ciphers to custom cryptographic techniques.
We evaluate state-of-the-art LLMs on CipherBank, e.g., GPT-4o, DeepSeek-V3, and cutting-edge reasoning-focused models such as o1 and DeepSeek-R1. Our results reveal significant gaps in reasoning abilities not only between general-purpose chat LLMs and reasoning-focused LLMs but also in the performance of current reasoning-focused models when applied to classical cryptographic decryption tasks, highlighting the challenges these models face in understanding and manipulating encrypted data. 
Through detailed analysis and error investigations, we provide several key observations that shed light on the limitations and potential improvement areas for LLMs in cryptographic reasoning.
These findings underscore the need for continuous advancements in LLM reasoning capabilities.

\end{abstract}

\begin{figure}[htbp]
    \centering
        \includegraphics[width=1\columnwidth]{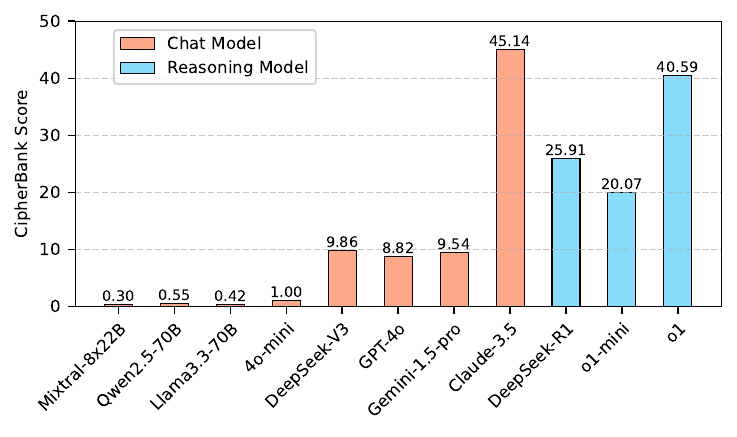}
        \caption{Comprehensive Performance of SOTA Chat and Reasoning Models on CipherBank.}
        \label{fig:subfig-a}
\end{figure}

\section{Introduction}

Large Language Models (LLMs) have revolutionized artificial intelligence by achieving state-of-the-art (SOTA) performance across diverse domains, from Natural Language Understanding (NLP)~\cite{dong2019unified,karanikolas2023large,sasaki2024enhancing} to complex problem-solving~\cite{yao2024hdflow,ge2023openagi}. Recent models, such as GPT-4o~\cite{hurst2024gpt} and Claude 3.5~\cite{anthropic2024claude}, have demonstrated unprecedented versatility, excelling in tasks ranging from creative writing to technical analysis. A particularly notable advancement lies in the reasoning-enhanced LLMs, which have emerged as a critical benchmark for evaluating LLMs' intelligence and now can solve mathematical problems~\cite{wu2024mathchat,ahn2024large,liu2024mathbench}, debug intricate code~\cite{lee2024unified,zhong2024ldb}, and even engage in multi-step logical deduction~\cite{sun2024determlr,wang2023can} with human-like proficiency.
For instance, specialized architectures like o1~\cite{openai2024o1} and DeepSeek-R1~\cite{guo2025deepseek} have pushed the boundaries of AI reasoning, achieving breakthroughs in domains such as theorem proving~\cite{yang2024leandojo} and algorithmic optimization~\cite{liu2024systematic}.
These achievements underscore the transformative potential of LLMs as general-purpose reasoning engines, capable of adapting to both broad and specialized challenges.

To quantify progress, the community has proposed numerous benchmarks targeting mathematical reasoning (e.g., MATH~\cite{hendrycksmath2021}, AIME\footnote{\url{https://huggingface.co/datasets/AI-MO/aimo-validation-aime}}, coding proficiency (e.g., HumanEval~\cite{chen2021codex}, MBPP~\cite{austin2021program}), and general logical deduction (e.g., FOLIO~\cite{han2024folionaturallanguagereasoning}, MMBench~\cite{MMBench}, CaLM~\cite{chen2024causal}. These testbeds have become indispensable tools for assessing model capabilities.

Despite extensive evaluations in mathematics and coding, one critical domain remains underexplored: cryptographic decryption. Cryptographic reasoning~\cite{shree2017review} demands unique capabilities, including pattern recognition, algorithmic Reverse-engineering, and contextual understanding of security constraints~\cite{schneier2002cryptographic}—skills distinct from those tested in conventional benchmarks. This gap is particularly consequential, as cryptography lies at the heart of modern digital security~\cite{konheim2007computer}, with applications spanning privacy-preserving communication~\cite{soomro2019review}, secure authentication~\cite{rani2022cyber}, and data integrity~\cite{sarkar2021role}. The absence of a rigorous benchmark for cryptographic reasoning not only limits the true understanding of LLM's reasoning ability but also hinders progress toward AI systems capable of contributing to security-critical contexts (e.g., jailbreaking~\cite{wei2024jailbroken}).
OpenAI has scratched the surface of this challenge and put a demo\footnote{\url{https://openai.com/index/learning-to-reason-with-llms/}} when releasing their strong reasoning model o1, but no serious efforts have been made to reveal this challenge in the committee. 

To address this gap, we introduce CipherBank, the first comprehensive benchmark specially designed to evaluate LLMs’ reasoning capabilities in cryptographic decryption tasks. CipherBank is meticulously constructed to reflect real-world scenarios requiring encryption, instead of general texts that may serve as a toy testbed, with 2,358 problems derived from 262 unique plaintexts across 5 domains (e.g., Personal Privacy, Financial Information) and 14 subdomains (e.g., Identity Information, Personal Income). 
As for cipher algorithms, it spans 3 major cryptographic categories—Substitution Ciphers (e.g., Rot13, Vigenère), Transposition Ciphers (e.g., Reverse, SwapPairs), and custom hybrid algorithms—encompassing 9 distinct encryption methods, covering \textbf{5} difficulty levels (from Basic to Expert) to ensure a diverse range of challenges. By integrating privacy-sensitive contexts and multi-layered cryptographic challenges, CipherBank provides a nuanced evaluation framework that captures both the complexity and practicality of real-world decryption tasks.

We evaluate CipherBank on SOTA LLMs, including general-purpose models (GPT-4o~\cite{hurst2024gpt}, DeepSeek-V3~\cite{liu2024deepseek}) and reasoning-optimized models (o1~\cite{openai2024o1}, DeepSeek-R1~\cite{guo2025deepseek}). Results reveal striking limitations: even advanced models struggle with classical ciphers, achieving only \textbf{$45.14$ }score on tasks solvable by human cryptanalysts. Notably, we observe a significant performance gap between general chat LLMs and specialized reasoning models, suggesting that current reasoning optimizations inadequately address cryptographic challenges. 
Besides, we also provide studies on different aspects for deep understandings, such as evaluate on noised plaintexts and different length of plaintexts. Observations show the limitations of current models in decryption reasoning, with chat and reasoning models each exhibiting distinct strengths and weaknesses in cryptographic tasks.
These findings highlight the need for targeted improvements in LLMs’ cryptographic reasoning, with implications for both AI safety (e.g., adversarial robustness) and applications in cybersecurity.

\section{CipherBank Construction}

CipherBank is a purpose-built benchmark designed to rigorously evaluate the reasoning capabilities of LLMs in cryptographic decryption tasks. It integrates three core components to ensure comprehensive coverage of real-world scenarios and cryptographic complexity: (1) \textbf{diverse plaintexts} meticulously constructed from multiple dimensions of real-world privacy-sensitive data, ensuring the decryption process aligns with practical requirements; (2) \textbf{a comprehensive suite of encryption algorithms}, including both traditional cryptographic methods and custom-designed algorithms, to thoroughly assess the model's reasoning, inductive, and computational capabilities from multiple perspectives; and (3) \textbf{a structured problem set with rich metadata}, enabling granular performance analysis and detailed error analysis based on the diverse properties of the plaintexts.

\begin{figure*}[htbp] 
    \centering
    \includegraphics[width=\textwidth]{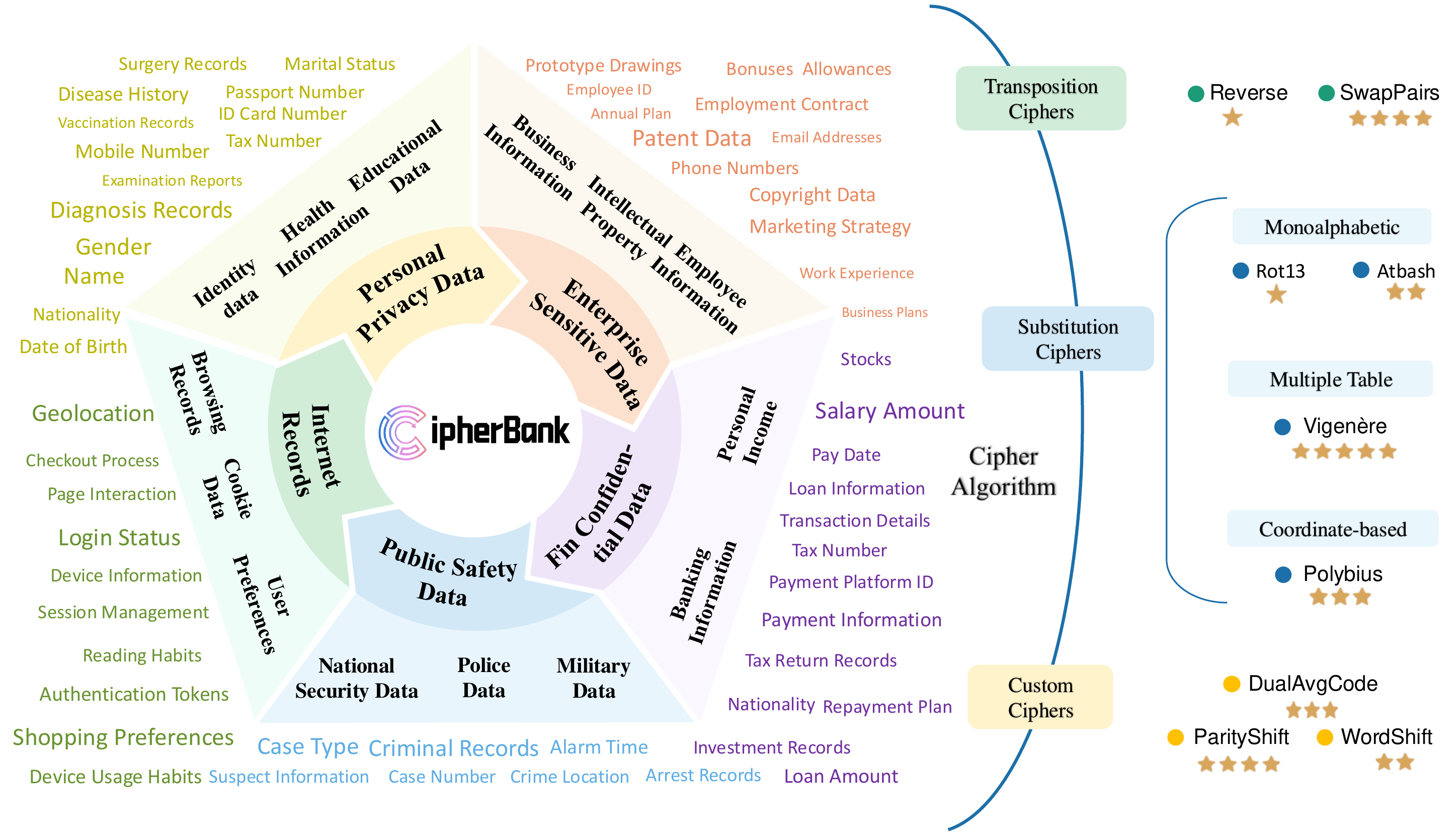} 
    \caption{Overview of CipherBank. CipherBank consists of simulated privacy data encrypted using various algorithms. The left side of the figure shows five domains, 14 subdomains, and selected tags. The right side displays three encryption categories, nine specific algorithms, and their corresponding difficulty levels.}
    \label{fig:pdf}
\end{figure*}

\subsection{Plaintext Data: Design, Sources, and Real-World Alignment} \label{sec:data}

To construct CipherBank, we meticulously analyze real-world encryption scenarios and categorize the corresponding data types into five primary domains: \textbf{Personal Privacy Data}, \textbf{Enterprise Sensitive Data}, \textbf{Public Safety Data}, \textbf{Financial Asset Data} and \textbf{Internet Records}. These domains are further refined into \textbf{14} subdomains (e.g., Health Information, Policy Data) to ensure comprehensive coverage of encryption needs. Inspired by UltraChat~\cite{ding2023enhancingchatlanguagemodels}, we adopt a \textbf{tag-based approach} to systematically structure encryption-relevant data, ensuring semantic consistency and domain relevance. Below, we detail the \textbf{3-step} process for generating high-quality plaintext data. 

\noindent{\textbf{Step 1:}} \textbf{Tag Definition and Curation.}
We leverage GPT-4o to generate candidate tags for each subdomain, capturing diverse real-world encryption scenarios. Human experts then curate these tags, eliminating redundancies, irrelevancies, and ambiguous entries, resulting in \textbf{89} distinct tags (see Appendix \ref{ref:tag}). This structured approach ensures that the generated plaintext data remains realistic, contextually meaningful, and representative of actual encryption use cases. The tags are designed to align with the \textbf{Variable Length property}, enabling the generation of inputs of varying sizes to assess model robustness.

\noindent{\textbf{Step 2:}} \textbf{Controlled Text Generation.}
Our plaintext generation process employs tag combinations to control text granularity: entries with more tags contain richer contextual details and greater length, while those with fewer tags remain concise and specific. To ensure semantic validity, all generated data are filtered to eliminate generic or redundant descriptions, creating a dataset that reflects diverse encryption scenarios with varying complexity. Additionally, we introduce the \textbf{Noise Perturbation} property through controlled noise injection, which serves two key objectives: (1) testing the model’s anti-interference capabilities and (2) reducing its reliance on contextual semantics to enhance robustness. Furthermore, we incorporate \textbf{Sensitive Numerical Data} by designing scenarios with complex alphanumeric combinations, including critical identifiers such as ID card and passport number. This multifaceted approach enables a comprehensive evaluation of the model’s ability to address sophisticated decryption challenges.

\noindent{\textbf{Step 3:}} \textbf{Expert Validation and Refinement.}
After generation, we conduct expert validation to ensure data quality, correctness, and relevance. Non-informative content, excessively long or short samples, and entries lacking clear privacy attributes are filtered out. Through this rigorous refinement process, we retain \textbf{262 }high-quality plaintext samples. This approach enables a practical and application-driven benchmark for evaluating LLMs’ decryption capabilities in cryptographic reasoning tasks.

\subsection{Encryption Algorithms} \label{sec:encryption}
CipherBank incorporates \textbf{3} major categories of encryption methods: Substitution Ciphers, Transposition Ciphers, and Custom Ciphers. \textbf{(1) Substitution-based techniques}, including Rot13, Atbash, Polybius and Vigenère, test a model’s ability to decode character-level transformations. These ciphers involve monoalphabetic or polyalphabetic substitutions, where each character is replaced by another based on a fixed rule or key. These methods evaluate the model’s capacity to decode symbolic mappings and generalize across substitution rules.
\textbf{(2) Transposition-based techniques}, such as Reverse and SwapPair, focus on positional rearrangements rather than symbol substitutions. These ciphers challenge the model to recognize structural patterns, such as reversed sequences or pairwise swaps. Unlike substitution ciphers, which alter character identities but preserve their order, transposition ciphers preserve characters but disrupt their sequence. This tests the model’s ability to analyze sequential dependencies and reconstruct the original symbol order.

To further assess LLMs’ ability to decrypt uncommon encryption methods, we introduce \textbf{(3) Custom-designed ciphers} that deviate from standard cryptographic schemes. \textbf{(a) DualAvgCode} is inspired by OpenAI’s o1 model showcase\footnote{\url{https://openai.com/index/learning-to-reason-with-llms/}}, where iterative transformations require models to infer multi-step encryption patterns. \textbf{(b) ParityShift} draws from LSB steganography~\cite{mielikainen2006lsb}, a common technique in information hiding, incorporating bitwise manipulations based on character parity. \textbf{(c) WordShift} Cipher is designed to evaluate LLMs’ ability to decrypt ciphers that combine substitution and transposition encryption, performing Caesar-style letter shifts within each word individually, blending character-level substitution with structural reordering.

Meanwhile, We categorize the nine algorithms into five difficulty tiers based on key necessity and computational complexity. T1 (Basic) includes simple ciphers like ROT13 and Reverse. T2 (Intermediate) introduces Atbash and WordShift with slightly more complex rules. T3 (Moderate) covers DualAvgCode and Polybius, requiring structured encoding. T4 (Advanced) involves ParityShift and SwapPairs with intricate data manipulation. T5 (Expert) features the Vigenère cipher, a polyalphabetic substitution cipher known for its keyword-based complexity. This framework organizes encryption techniques from basic to expert.

\begin{figure*}[htbp]
    \centering
    \begin{exmp}{Plain-Ciphertext Pair}{parallel}
        \small
        \setstretch{1.4} 

        \# Domain: Personal Privacy Data

        \#\# Subdomain: Identity Information

        \#\#\# Tag Combination: ["Name", "Date of Birth", "Passport Number"]  

        \textbf{Plaintext}:
        
        \textit{Peter was born on April 23, 1985, and carries a passport with the number X123456789. } 

        \textbf{Encryption results:}
        
        (1) \textbf{Rot13}:\textit{ Crgre jnf obea ba Ncevy 23, 1985, naq pneevrf n cnffcbeg jvgu gur ahzore K123456789.  }

        (2) \textbf{SwapPairs}: \textit{ePet raw sobnro  npAir l32 ,9158 ,na dacrrei s aapssoptrw ti hht eunbmreX 21436587.9 } 

        (3) \textbf{WordShift} : \textit{erPet was nbor no ilApr 23, 5,198 and riescar a sportpas hwit the bernum 3456789.X12  }

        (4) ...
        
         More results can be found in the appendix.
    \end{exmp}
\end{figure*}

\subsection{CipherBank Statistics}

As shown in Figure~\ref{fig:pdf}, we provides an overview of CipherBank structure. The encryption algorithm in Section~\ref{sec:encryption} applies to the expert-curated dataset from Section~\ref{sec:data}, yielding \textbf{2,358} test data points.

\begin{table}[htbp]
    \centering
    \caption{Statistics of CipherBank.}
    \label{table1}
    \resizebox{\columnwidth}{!}{
    \begin{tabular}{lcccc}
        \toprule
        \textbf{Domains}  & \textbf{\#Tag} & \textbf{\#Plaintext}  & \textbf{\#Test} & \textbf{Avg(len)} \\
        \midrule
        Personal Privacy Data & 23 &50 & 450&107.88 \\ 
        Enterprise Sensitive Data & 16 & 52 & 468&103.10 \\ 
        Public Safety Data & 17& 63 & 567&110.89  \\ 
        Financial Asset Data & 13& 44 & 396 &163.68  \\
        Internet Records &20& 53 & 477 & 191.92 \\
        \midrule
        Summary & 89 & 262 & 2358 & 134.03 \\ \midrule
        \bottomrule
    \end{tabular}
    }
\end{table}

Table~\ref{table1} summarizes the distribution of plaintexts across 5 domains, each with varying numbers of tags, samples, and test cases. Notably, Internet Records has the longest plaintexts (191.92), while Enterprise Sensitive Data has shorter samples (103.10). This diversity ensures a comprehensive evaluation of model performance across different encryption contexts.

\section{Evaluations}
\subsection{Evaluation Setup}

\noindent{\textbf{Evaluation Protocols.}}  
In terms of testing methodology, CipherBank’s evaluation follows the Known-Plaintext Attack framework \cite{zulkifli2008attack}, employing a \textbf{3-shot} testing approach. We prompt the model with three plaintext-ciphertext pairs as demonstrations to infer encryption rules, identify potential keys, and apply the learned patterns to decrypt a new ciphertext. The detailed prompt can be found in Appendix~\ref{sec:prompt}.

For evaluation metrics, we primarily employ \textit{accuracy} to measure overall decryption success, which is the ratio of correctly decrypted cases to total test cases, where correctness requires an exact character match with the plaintext.
Additionally, to capture finer-grained differences between the decrypted output and the original plaintext, we incorporate \textit{ Levenshtein similarity  }\cite{yujian2007normalized}. We compute the Levenshtein distance for each sentence individually and report the average Levenshtein similarity across all test cases, providing a more nuanced assessment of model performance beyond binary correctness.

\noindent{\textbf{LLM Candidates.}}  
For a comprehensive evaluation, we carefully selected \textbf{18} SOTA LLMs for evaluation, ensuring a diverse representation of open-source, closed-source, and reasoning-specialized models. Below, we outline the tested models:  

\noindent{\textbf{$\star$ \textit{Open-Source Chat Models:}}} We evaluate leading open-source LLMs, including Mistral AI's Mixtral-8x22B \cite{jiang2024mixtral}, Alibaba's Qwen2.5-72B-Instruct \cite{yang2024qwen2}, Meta's Llama-3.1-70B-Instruct and Llama-3.3-70B-Instruct \cite{dubey2024llama}, as well as the rising star - DeepSeek-V3 \cite{liu2024deepseek}.  

\noindent{\textbf{$\star$ \textit{Closed-Source Models:}}} For proprietary models, evaluation is conducted via API access. The tested models include OpenAI's 4o-mini and GPT-4o series (0806, 1120) \cite{hurst2024gpt}, DeepMind's Gemini-1.5-Pro \cite{geminiteam2024gemini15unlockingmultimodal} and Gemini-2.0-Flash-Exp\footnote{\url{https://deepmind.google/technologies/gemini/flash/}}, along with Anthropic's Claude-Sonnet-3.5 (1022)\footnote{\url{https://www.anthropic.com/news/claude-3-5-sonnet}}.  

\noindent{\textbf{$\star$ \textit{Reasoning Models:}}} We further investigate models optimized for reasoning tasks, including QwQ-32B-Preview \cite{qwq-32b-preview}, DeepSeek-R1 \cite{guo2025deepseek}, Gemini-2.0-Flash-Thinking (1219)\footnote{\url{https://deepmind.google/technologies/gemini/flash-thinking/}}, o1-mini (0912) and o1 (1217) \cite{openai2024o1}.  

\begin{table*}[h!]
    \centering
    \caption{3-shot scores (\%) of LLMs across three major encryption paradigms and nine specific encryption algorithms on CipherBank. The highest scores in each category are highlighted with a \colorbox[HTML]{d6e7f5}{\textbf{blue background}}, while the second-best results are \underline{underlined} for emphasis.}
    \label{table2}
    \resizebox{\textwidth}{!}
    {
    \begin{tabular}{lcccccccccccc}
        \toprule
        \multirow{2}{*}{\textbf{Model}} & \multicolumn{4}{c}{\textbf{Substitution Ciphers}} & \multicolumn{2}{c}{\textbf{Transposition Ciphers}}  & \multicolumn{3}{c}{\textbf{Custom Ciphers}}  & \multicolumn{1}{c}{\textbf{Cipher Score}}  \\
        \cmidrule(lr){2-5} \cmidrule(lr){6-7} \cmidrule(lr){8-10} \cmidrule(lr){11-11} 
        &Rot & Atbash & Polybius & Vigenère & Reverse & SwapPairs  & DualAvgCode & ParityShift & WordShift & Accuracy$_{\text{avg}}$\\
        \midrule
        \rowcolor[HTML]{f8e2d1} 
        \multicolumn{11}{c}{\textit{Open-source Chat Models}} \\ \midrule
        Mixtral-8x22B-v0.1   & 0.38  & 0 & 0 & 0 & 0.76  & 0   & 0.38  & 0 &1.15 & 0.30 \\
        Qwen2.5-72B-Instruct & 1.15  & 0 & 0 & 0 &0 & 0.38  &1.15&0 &2.29 & 0.55\\ 
        Llama-3.1-70B-Instruct  & 1.15 & 0.38 & 0 & 0.38 & 0 & 0  & 0.38 & 0.38 &0.76 & 0.38\\ 
        Llama-3.3-70B-Instruct & 2.67 & 0.38 & 0 & 0 & 0 & 0 & 0 & 0.76 & 0 & 0.42\\ 
        DeepSeek-V3 & 32.44 &  14.88 & 2.29 & 0.76& 28.47 & 0.38   &  0.38& 1.14 & 8.02 & 9.86\\  \midrule
        \rowcolor[HTML]{fdf3d0} 
        \multicolumn{11}{c}{\textit{Closed-source Models}} \\ \midrule
        GPT-4o-mini-2024-07-18 & 3.69& 2.03 & 0 & 0.51 & 2.16 & 0  & 0.38 & 0 & 0.25 & 1.00 \\
        GPT-4o-2024-08-06 & 38.17 & 3.05&0.38&0.76&25.19&2.29&0&1.14&8.40 & 8.82\\
        GPT-4o-2024-11-20 & 26.46& 6.99 & 0.13 &0.76 & 15.27 & 0.76  &0.25 &  0.89 &  6.11&6.40\\
        gemini-1.5-pro & 55.34& 0.76 & 0.38 &0.76 & 10.31 & 0.76 &0.38 &  0.76 &  16.41 & 9.54\\
        gemini-2.0-flash-exp & 35.88 & 3.05& 1.53 & 0.38&\underline{29.39} &1.53  & 0 &0.76&5.34 & 8.65\\
        Claude-Sonnet-3.5-1022 &\cellcolor[HTML]{d6e7f5}\textbf{83.21} & \underline{75.19} & \underline{72.90} & \underline{1.91} & \cellcolor[HTML]{d6e7f5}\textbf{63.93} & \underline{6.87}   &  4.96 & \cellcolor[HTML]{d6e7f5}\textbf{58.21} &  \cellcolor[HTML]{d6e7f5}\textbf{39.12} & \cellcolor[HTML]{d6e7f5}\textbf{45.14}   \\
        \midrule
        \rowcolor[HTML]{d6ecda} 
        \multicolumn{11}{c}{\textit{ Reasoning Models}} \\ \midrule
        QwQ-32B-Preview  & 1.53 & 0.38 & 1.91 & 0 & 0 & 0 &  0.38 & 0.38 & 2.29 & 0.76\\
        DeepSeek-R1  & \underline{73.28} & 58.78&44.27& 0.38&10.69& 0.38 & \underline{24.05} & \underline{12.98} &8.40 & 25.91\\ 
        gemini-2.0-flash-thinking  & 40.46  &17.18 &21.76 & 1.15 & 22.90& 1.15  & 0 & 7.63 & 9.16 &13.49  \\
        o1-mini-2024-09-12 & 46.18 & 68.32 & 46.95 &1.53 & 5.15 & 0.38  &2.93& 7.63 & 1.53 &  20.07\\
        o1-2024-12-17  & 59.92  &\cellcolor[HTML]{d6e7f5}\textbf{79.01} & \cellcolor[HTML]{d6e7f5}\textbf{79.39} &\cellcolor[HTML]{d6e7f5}\textbf{7.25} &14.89 & \cellcolor[HTML]{d6e7f5}\textbf{32.14} &\cellcolor[HTML]{d6e7f5}\textbf{50.38} & 12.39 & \underline{29.90} & \underline{40.59}\\
        \midrule
        \bottomrule
    \end{tabular}%
    
    }
\end{table*}

\subsection{Benchmark Results}
Table~\ref{table2} presents the evaluation results of all candidate LLMs (Levenshtein similarity results are in Appendix \ref{LD_result}). 
Below, we distill the experimental findings into several observations:

\noindent{\textbf{Limitations of Current Models in Cryptographic Reasoning.}}  
Despite advancements in LLMs, Table~\ref{table2} highlights their limitations in structured cryptographic reasoning. The overall performance remains low, with most SOTA models struggling to achieve meaningful accuracy. In Cipher Score, common models like Qwen and LLaMA perform particularly poorly, with some scoring in the single digits or near zero. Even the best-performing models, Claude-3.5 and o1, achieve less than 50 in accuracy, underscoring the significant difficulty of CipherBank and the challenges LLMs face in systematic decryption.

\noindent{\textbf{Reasoning Models Generally Outperform Chat Models.}}
When comparing reasoning models to chat models, generally we can find that the reasoning models do outperform chat models on all cipher algorithms and achieve better overall performance. The only expectation is the superior performance of Claude-3.5 (45.14) even better than o1, and also the bad performance of QwQ-32B-Preview (only 0.76 accuracy). This clearly demonstrate the advantages of the reasoning-specialized models. 

\noindent{\textbf{Closed-Source Models Retain an Edge Over Open-Source Models.}}
Overall, closed-source models outperform open-source models in cryptographic decryption. Claude-3.5 (45.14) and o1 (40.59) achieve the highest performance across all cipher categories. However, DeepSeek-V3 (9.86) and DeepSeek-R1 (25.91) surpass most models in the GPT and Gemini families, indicating that advanced open-source models are closing the gap. Nevertheless, both still lag behind Claude-3.5 and o1, suggesting that while open-source models are improving, there is significant potential for open-source models to achieve even better performance in the future.

\noindent{\textbf{The performance variance among models of the same category is remarkably significant.}}
 Within the Open-source Chat Models category, the top-performing model, deepseek-v3 (9.86), outperforms the weakest model, Mixtral-8x22B (0.30), by a factor of 33. Similarly, in the Closed-source Models category, Claude-Sonnet-3.5 (45.14) demonstrates a performance 45 times greater than that of GPT-4o-mini (1.00). The disparity is even more pronounced in the Reasoning Models category, where o1 (40.59) surpasses QwQ-32B-Preview (0.76) by a factor of 53. Such substantial performance variations are rarely observed in other benchmarks, highlighting the challenging nature of CipherBank. This benchmark effectively distinguishes the reasoning capabilities of different models through its decryption dimension, providing a robust framework for evaluating model performance.

\begin{figure*}[t] 
    \centering
    \begin{subfigure}[b]{0.32\textwidth} 
        \includegraphics[width=\textwidth]{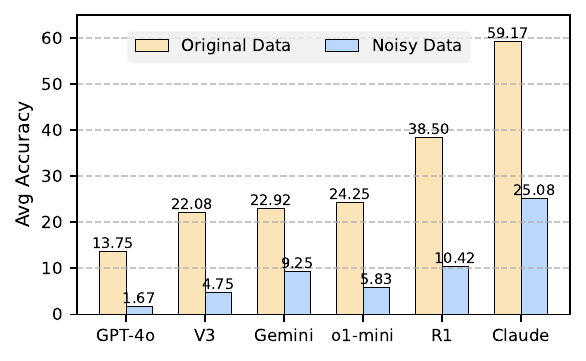}
        \caption{Model Robustness to Noisy Inputs: Performance Comparison.}
        \label{fig:1}
    \end{subfigure}
    \hfill 
    \begin{subfigure}[b]{0.32\textwidth}
        \includegraphics[width=\textwidth]{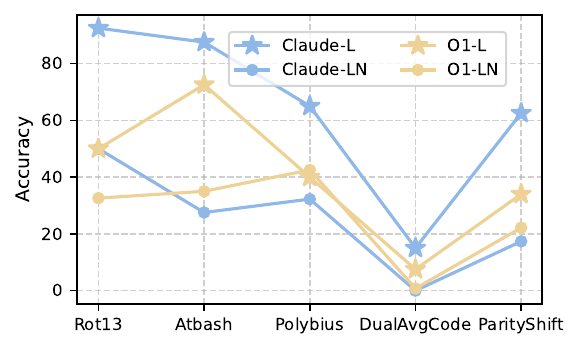}
        \caption{Effect of Encryption Scope: Letters Only vs. Letters \& Numbers.}
        \label{fig:2}
    \end{subfigure}
    \hfill 
    \begin{subfigure}[b]{0.32\textwidth}
        \includegraphics[width=\textwidth]{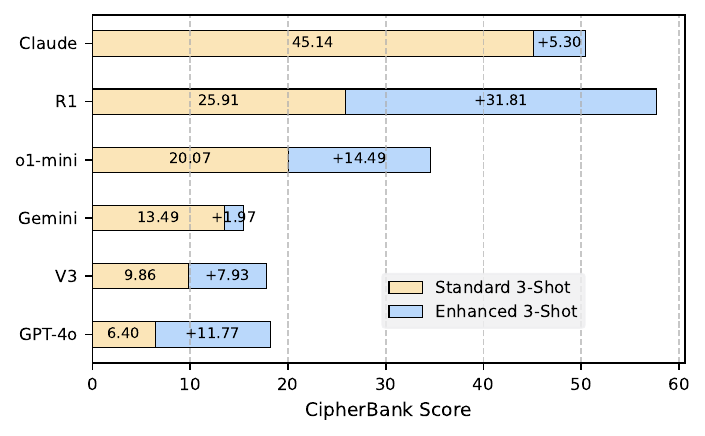}
        \caption{Evaluating the Benefit of Explicit Algorithm Hints in 3-Shot Prompting.}
        \label{fig:3}
    \end{subfigure}
    \caption{Evaluation of LLM Performance Under Different Encryption and Prompting Conditions.}
    \label{fig:all}
\end{figure*}

\section{Detailed Analysis}
In this section, we conduct a detailed analysis from the perspectives of plaintext characteristics, noise levels, testing methodologies, finer-grained evaluation metrics, and error analysis to gain deeper insights into the strengths and limitations of different LLMs in cryptographic decryption.

\begin{table}[h!]
    \centering
    \caption{Model Performance on Short and Long Plaintext Setting (Lower Difference and Decrease Ratio Are Better). We highlight the most stable and sensitive results in \colorbox[HTML]{d6e7f5}{blue} and  \colorbox[HTML]{d6ecda}{green} respectively.}
    \label{ana_1}
    \resizebox{\columnwidth}{!}{%
    \begin{tabular}{lcccc}
        \toprule
        \textbf{Model} & \textbf{Short} & \textbf{Long} & \textbf{Diff} & \textbf{Decrease Ratio(\%)} \\
        \midrule
        GPT-4o-2024-11-20 & \cellcolor[HTML]{d6ecda}9.47 & \cellcolor[HTML]{d6ecda}4.46 & 5.01 & 52.60 \\
        gemini-2.0-flash-exp & 11.50 & 6.42 & 5.08 & 44.35 \\
        DeepSeek-V3 & 13.24 & 5.22 & 8.02 & \cellcolor[HTML]{d6ecda}60.60 \\
        gemini-2.0-flash-thinking & 19.90 & 8.47 & 11.43 & 42.61 \\
        DeepSeek-R1 & 32.27 & 20.94 & 11.33 & 33.16 \\
        o1-mini-2024-09-12 & 33.77 & 17.35 & \cellcolor[HTML]{d6ecda}16.42 & 48.57 \\
        o1-2024-12-17 & 47.61 & 34.38 & 13.23 & 27.78 \\
        Claude-Sonnet-3.5 & \cellcolor[HTML]{d6e7f5}48.70 & \cellcolor[HTML]{d6e7f5}47.85 & \cellcolor[HTML]{d6e7f5}0.85 & \cellcolor[HTML]{d6e7f5}1.74 \\ \midrule
        \bottomrule
    \end{tabular}
    }
\end{table}

\subsection{Impact of Plaintext Length}
To test models’ sensitivity to text length, we categorize plaintexts into short (fewer than three tags) and long groups, averaging $70.29$ and $181.61$ characters, respectively. As shown in Table \ref{ana_1} (full results and plaintext examples can be found in Appendix~\ref{appendix_analysis}), longer plaintexts lead to a significant performance decline in most models.
Most models exhibit a significant decline in decryption performance as text length increases. Among them, Claude-3.5 (-0.85) shows the most stable performance, while o1-mini (-16.42) is the most sensitive. This contrasts with human performance, highlighting LLMs’ \textbf{length bias} in decryption reasoning.
\subsection{Effect of Noise on Model Robustness}
We observe that models frequently substituted synonyms instead of strictly applying decryption rules to each character (examples in Appendix~\ref{appendix_analysis}), indicating the presence of \textbf{shortcut reasoning}, where models partially decrypt the text and infer the remainder based on semantic context rather than adhering to the encryption pattern.

To evaluate robustness and mitigate reliance on semantic inference, we select the 40 plaintexts with the lowest \textbf{perplexity} (PPL) scores, computed using Llama-3.1-8B-Instruct, for noise injection. Figure \ref{fig:1} shows a substantial performance drop across all models, including Claude-3.5 (from 59.17 to 25.08) and o1-mini (from 24.25 to 5.83), highlighting their vulnerability to structural perturbations and further exposing the limitations of current models in systematic reasoning and precise decryption.

\subsection{Effect of Encryption Scope}

In previous evaluations, only letters are encrypted. To better reflect real-world scenarios, here we select plaintexts with sensitive numerical data and apply encryption to both letters and numbers, focusing on algorithms that directly affect numbers (test prompt in Appendix~\ref{appendix_analysis}).  
As shown in Table \ref{fig:2}, model performance drops significantly in this more complex setting. This suggests difficulty in adapting decryption strategies to numerical transformations. Even under the same encryption principles, encrypting both letters and numbers greatly increases task complexity, posing a significant challenge for current reasoning models. This highlights a critical limitation in LLMs’ ability to generalize across diverse data types, particularly when numerical transformations are involved. Future work should focus on enhancing models’ capacity to handle mixed data encryption.

\subsection{Effect of Explicit Algorithm Hints on Decryption Performance}

Previous evaluations highlight the significant challenges posed by CipherBank. To evaluate the models’ decryption capabilities when provided with algorithm details, we enhance the 3-shot setting by explicitly informing the models of the specific algorithm during testing. Under the revised setting, models are no longer required to independently deduce encryption logic but instead focus on identifying the necessary key and applying the specified decryption rules. The enhanced prompt is provided in Appendix~\ref{appendix_analysis}. Table~\ref{fig:3} reveals distinct performance patterns. Most chat models show minimal improvement even with algorithm details, struggling with key inference and decryption—highlighting persistent limitations, especially in models like Claude (+5.30) and Gemini (+1.97).

In contrast, reasoning models show marked performance gains, with R1 (+31.81) and o1-mini (+14.49) achieving significant improvements.  
The observed contrast underscores a fundamental distinction: chat models primarily rely on surface-level pattern recognition, while reasoning models excel in structured inference when provided with appropriate guidance.
\subsection{Error Analysis}

We conduct a comprehensive error analysis based on the test results in Table~\ref{table2}, identifying six distinct error types. To gain deeper insights, we examine the three best-performing chat models and three best-performing reasoning models, summarizing their error distributions. Detailed error definitions and examples are provided in Appendix \ref{error1} and \ref{error2}.

\begin{figure}[htbp] 
    \centering
    \includegraphics[width=\columnwidth]{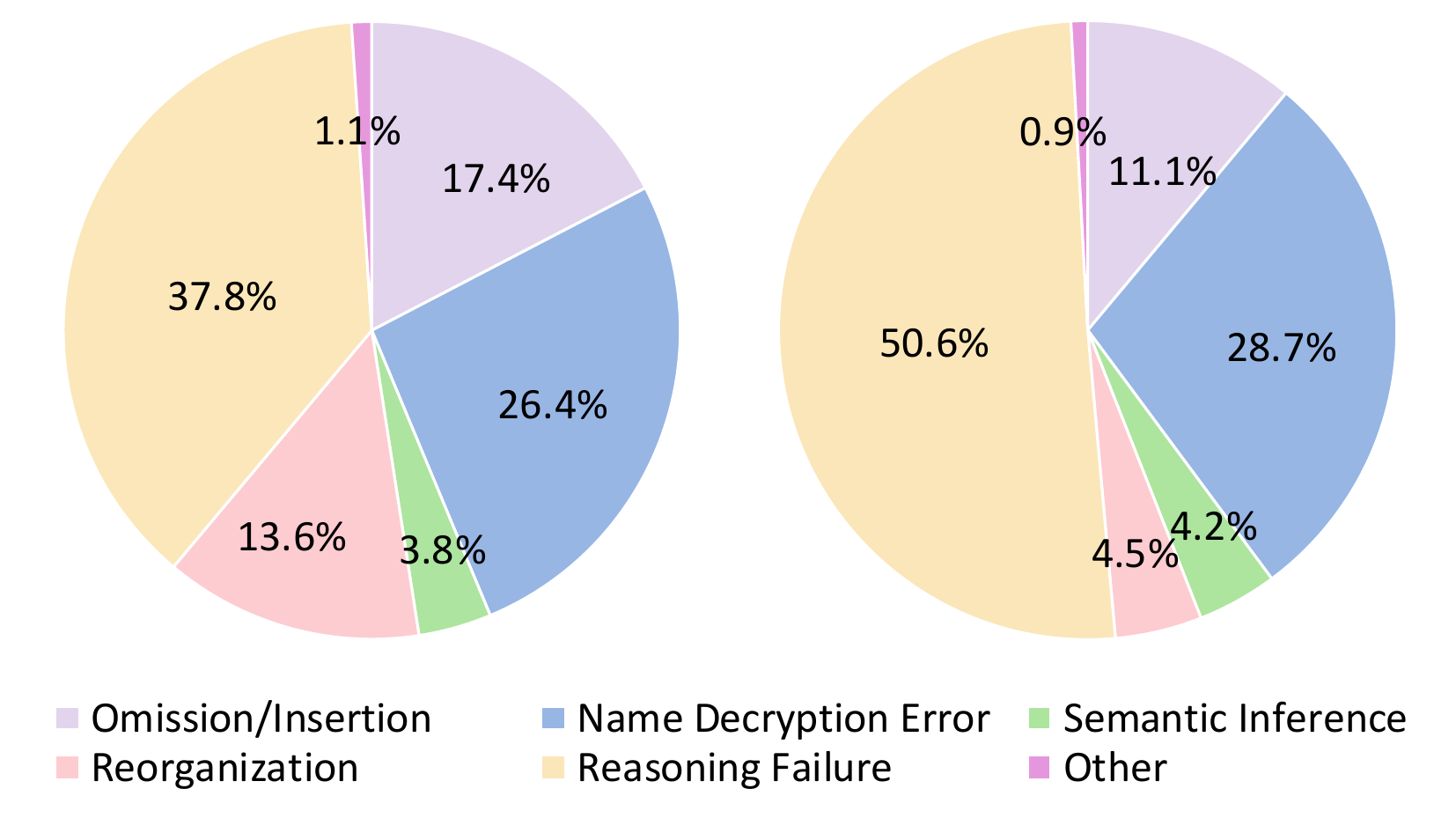} 
    \caption{Decryption Error Distribution. The left represents chat models, while the right corresponds to reasoning models.}
    \vspace{-0.6cm}
    \label{fig:error}
\end{figure}

As shown in Figure \ref{fig:error}, the distribution of error types reveals key differences between reasoning and chat models.
Surprisingly, (1) reasoning models exhibit a higher rate of reasoning failures than chat models. A deeper examination of Appendix \ref{error3} reveals that many of these failures occur on simpler tasks, suggesting that reasoning models may overanalyze problems, leading to incorrect conclusions. This indicates that their complex inference processes can sometimes hinder performance on straightforward decryption cases.
Conversely, (2) chat models show a higher frequency of omission/insertion and reorganization errors, indicating that while they are stronger in semantic understanding, this often results in excessive auto-completion and sentence restructuring rather than strict rule adherence. This tendency suggests that chat models prioritize fluency over exact decryption, leading to unintended modifications.
Additionally, (3) both model types frequently make errors in name decryption, highlighting a broader challenge in handling structured entity transformations. This suggests that current LLMs struggle to consistently apply encryption rules to proper nouns, potentially due to memorization biases or difficulties in preserving entity-level consistency during decryption.

\section{Related Work}

\paragraph{Benchmarks for Reasoning}

Evaluating reasoning abilities in LLMs has been a key focus in AI research, with various benchmarks assessing models across mathematical, logical, and inferential tasks. MATH~\cite{hendrycks2021measuring}, MathBench~\cite{liu2024mathbench}, and LiveMathBench~\cite{liu2024your} test arithmetic and algebraic reasoning, while HumanEval~\cite{chen2021evaluating}, DebugBench~\cite{tian2024debugbench} and BigCodeBench~\cite{zhuo2024bigcodebench} evaluates code generation that require programming logic. Additionally, BIG-Bench~\cite{srivastava2022beyond}, BBH~\cite{suzgun2022challenging}, and LiveBench~\cite{white2024livebench} measure broader cognitive abilities, such as abstract reasoning and analogical problem-solving. KOR-Bench~\cite{ma2024kor} is new benchmark that examines strong reasoning by introducing Knowledge-Orthogonal Reasoning (KOR) tasks, assessing models’ ability to apply newly introduced rules independent of pretrained knowledge. Specially, it also contains a cipher reasoning task, which provides explicit encryption rules and keys, guiding models through step-by-step decryption rather than requiring pattern inference.
In contrast, CipherBank presents a more realistic challenge, requiring models to identify encryption patterns from examples without prior knowledge, better reflecting real-world scenarios where encryption schemes are unknown.

\paragraph{Jailbreaking via Cipher Characters}

Recent work demonstrates that encoding adversarial prompts via encryption~\cite{yuan2023gpt,wei2024jailbroken} or obfuscation~\cite{yong2023low,jiang2024artprompt,kang2024exploiting} can bypass LLM safety filters by exploiting models’ ability to process encoded inputs. While CipherBench~\cite{handa2024competencyreasoningopensdoor} evaluates cipher-based jailbreaking, its reliance on 40 curated plaintexts and explicit algorithm hints limits practical relevance. Our CipherBank removes prior guidance, requiring autonomous pattern inference from plaintext-ciphertext pairs to simulate privacy-sensitive decryption scenarios, establishing a robust benchmark for LLM security evaluation.

\section{Conclusion}
In this work, we introduce CipherBank, a comprehensive benchmark for evaluating reasoning capabilities through cryptographic decryption. CipherBank includes 5 domains, 14 subdomains of plaintext data, 9 encryption algorithms, and 2,358 decryption tasks. By testing SOTA LLMs on CipherBank, we uncover significant limitations in their decryption abilities, revealing distinct strengths and weaknesses between reasoning and chat models. Our analysis identifies key deficiencies in current reasoning approaches and suggests directions for improvement, positioning CipherBank as a novel benchmark for advancing structured inference and cryptographic reasoning in developing future LLMs.

\section*{Limitations}

Our evaluation is constrained by the reliance on closed-source models, which are accessible only via API calls. This introduces potential variability due to API updates and version changes, though we mitigate this by documenting the specific versions and dates used. Additionally, access restrictions prevent us from evaluating more advanced models such as o1 Pro and o3 series, limiting the scope of our benchmark. From a design perspective, CipherBank primarily focuses on classical encryption algorithms, as modern cryptographic schemes introduce complexities beyond current model capabilities. While this choice ensures feasibility in evaluation, it also restricts the benchmark’s applicability to real-world cryptographic challenges. As models improve, expanding CipherBank to modern encryption techniques will provide a more comprehensive assessment of reasoning in cryptographic tasks.

\bibliography{custom}

\appendix

\newpage
\onecolumn
\twocolumn

\section{Detailed Benchmark Description}
In this chapter, we provide additional details on CipherBank that were not extensively covered in the main text. This includes a detailed breakdown of plaintext tags and their distribution across subdomains, as well as a more comprehensive description of the encryption algorithms used. These details offer deeper insights into the dataset construction and the encryption schemes evaluated in this benchmark.
\subsection{Tags and Plaintext Distribution Across Subdomains}
\label{ref:tag}

\begin{table*}[p]
    \centering
    \renewcommand{\arraystretch}{1.5} 
    \caption{Tag Distribution Across Subdomains in CipherBank}
    \label{tab:subdomain_tags}
    \resizebox{\textwidth}{!}{%
    \begin{tabular}{|m{4cm}|m{4cm}|p{12cm}|}  
        \hline
        \textbf{Domain} & \textbf{Subdomain} & \textbf{Tags} \\
        \hline
        \multirow{3}{=}{\centering Personal Privacy Data} 
        & \centering Identity Information & Name, ID Card Number, Passport Number, Date of Birth, Gender, Nationality, Marital Status, Mobile Number, Family Member Information (e.g., immediate family names, contact information), Residential Address \\
        \cline{2-3}
        & \centering Health Information & Medical Record Number (Patient ID), Diagnosis Records, Surgery Records, Examination Reports (e.g., X-ray, CT scan results, heart rate, blood pressure, blood sugar level, blood type), Disease History, Allergy History, Vaccination Records, Family Medical History \\
        \cline{2-3}
        & \centering Educational Data & Student ID (Student Number), School Records (Enrollment Date, Graduation Date), Academic Records (Subjects, Grades, GPA, Ranking), Degree Information (Bachelor, Master, Doctorate), Awards and Penalties Records (Disciplinary Records) \\
        \hline
        \multirow{3}{=}{\centering Enterprise Sensitive Data} 
        & \centering Business Information & Business Plans (e.g., Annual Plan, Five-Year Plan), Marketing Strategy (e.g., Marketing Promotion Plan, Advertising Budget), Customer Lists (e.g., Customer Contacts, Preferences), Supplier Information (Supplier List, Cooperation Agreements), Internal Financial Budgets (Cost Structure, Profit Forecasts) \\
        \cline{2-3}
        & \centering Intellectual Property & Product Design Plans (e.g., Prototype Drawings, Design Documents), Internal Technical Documents (e.g., Technical Manuals, Specifications), Test Data (e.g., Product Performance Test Results, Quality Control Records), Copyright Data, Patent Data \\
        \cline{2-3}
        & \centering Employee Information & Contact Information (e.g., Phone Numbers, Email Addresses), Work Experience, Position and Department Information, Salary and Benefits Information (e.g., Salary Amount, Bonuses, Allowances), Performance Evaluation (e.g., Performance Scores, Promotion Records), Contract Information (e.g., Employment Contract, Non-Disclosure Agreement) \\
        \hline
        \multirow{3}{=}{\centering Public Safety Data} 
        & \centering Police Data & Case Information (Case Number, Case Type, Filing Date), Criminal Records (Suspect Information, Crime Time, Crime Location), Alarm Records (Informer Information, Alarm Time, Alarm Content), Investigation Reports (Investigation Results, Investigation Progress), Arrest Records (Arrest Time, Location, Action Description), Traffic Enforcement Data (Violation Records, Penalty Information), Police Officer Information (Officer Number, Name, Position, Department), Police Resource Allocation (Vehicle, Equipment, Weapon Usage Records) \\
        \cline{2-3}
        & \centering National Security Data & Border Crossing Records (Entry and Exit Personnel Information, Vehicle Registration), Customs Inspection Data (Cargo List, Contraband Records), Territorial Patrol Data (Patrol Reports, Anomalies Records), Cyber Security Monitoring Data (Cyber Attack Records, Threat Intelligence) \\
        \cline{2-3}
        & \centering Military Data & Operation Plans, Target Location, Troop Deployment, Military Base Distribution, Defense Works Location \\
        \hline
        \multirow{2}{=}{\centering Financial Confidential Data} 
        & \centering Banking Information & Account Number, Bank Card Number, Payment Method, Payment Platform ID, Transaction Details, Loan Amount, Interest Rate, Repayment Plan, Investment Records (Stocks, Funds, Bonds) \\
        \cline{2-3}
        & \centering Personal Income & Salary Amount, Pay Date, Tax Number, Tax Return Records \\
        \hline
        \multirow{3}{=}{\centering Internet Records} 
        & \centering Browsing Records & Page Interaction, Search Behavior, Click Activity, Device Information, Geolocation, Checkout Process, Multimedia Interaction, Download Records \\
        \cline{2-3}
        & \centering Cookie Data & Session Management, User Identification, Ad Targeting, Behavior Tracking, Authentication Tokens, Login Status \\
        \cline{2-3}
        & \centering User Preferences & Preferred Genres, Device Usage Habits, Notification Preferences, Shopping Preferences, Video Preferences, Reading Habits \\
        \hline
    \end{tabular}
    }
\end{table*}

\textbf{Table~\ref{tab:subdomain_tags}} provides an overview of the specific tags associated with each subdomain within CipherBank. The dataset spans \textbf{five primary domains} and \textbf{14 subdomains}, ensuring diverse and realistic plaintext scenarios for cryptographic evaluation.

\begin{figure*}[p]
    \centering
    \begin{exmp}{Plain-Ciphertext Pair (Identity Information) - Only Letter}{parallel}
        \small
        \setstretch{1.4}  

        \# Domain: Personal Privacy Data

        \#\# Subdomain: Identity Information

        \#\#\# Tag Combination: ["Name", "Date of Birth", "Passport Number"]  

        \textbf{Plaintext}:
        
        \textit{Peter was born on April 23, 1985, and carries a passport with the number X123456789.} 

        \textbf{Encryption results:}
        
        (1) \textbf{Rot13}: \textit{Crgre jnf obea ba Ncevy 23, 1985, naq pneevrf n cnffcbeg jvgu gur ahzore K123456789.}

        (2) \textbf{Atbash}: \textit{Kvgvi dzh ylim lm Zkiro 23, 1985, zmw xziirvh z kzhhklig drgs gsv mfnyvi C123456789.}

        (3) \textbf{Polybius}: \textit{34 15 42 15 36   45 11 41   12 33 36 32   33 32   11 34 36 23 26   2 3 ,   1 9 8 5 ,   
        11 32 14   13 11 36 36 23 15 41   11   34 11 41 41 34 33 36 42   45 23 42 22   42 22 15   32 43 31 12 15 36   
        46 1 2 3 4 5 6 7 8 9.}

        (4) \textbf{Vigenère}: \textit{Pgeet wcd dzrp op Arcin 23, 1985, cyd natcigd c pcdsrzrv wkeh ehg nwxbgc Z123456789.}

        (5) \textbf{Reverse}: \textit{.987654321X rebmun eht htiw tropssap a seirrac dna ,5891 ,32 lirpA no nrob saw reteP}

        (6) \textbf{SwapPairs}: \textit{ePet raw sobnro  npAir l32 ,9158 ,na dacrrei s aapssoptrw ti hht eunbmreX 21436587.9}

        (7) \textbf{DualAvgCode}: \textit{OQdfsudfqs vxaart acnpqsmo npmo AAoqqshjkm 23, 1985, aamoce bdaaqsqshjdfrt aa oqaartrtoqnpqssu vxhjsugi sugidf motvlnacdfqs WY123456789.}

        (8) \textbf{ParityShift}: \textit{Qduds vzr cnso no Zqshm 23, 1985, zoe bzsshdr z qzrrqnsu vhui uid otlcds Y123456789.}

        (9) \textbf{WordShift}: \textit{erPet was nbor no ilApr 23, 5,198 and riescar a sportpas hwit the bernum 3456789.X12}

    \end{exmp}
\end{figure*}

\begin{figure*}[p]
    \centering
    \begin{exmp}{Plain-Ciphertext Pair (Police Data) - Only Letter}{parallel}
        \small
        \setstretch{1.4}  

        \# Domain: Public Safety Data

        \#\# Subdomain: Police Data

        \#\#\# Tag Combination: ["Suspect Information", "Crime Time", "Crime Location", "Police Officer Information"]  

        \textbf{Plaintext}:
        
        \textit{Suspect: Jonathan, Crime: Burglary, Time: 2022-03-12 14:30, Location: 123 Elm Street, Officer Smith observed suspicious activity near 5th Ave on 2022-03-13.} 

        \textbf{Encryption results:}
        
        (1) \textbf{Rot13}: \textit{Fhfcrpg: Wbanguna, Pevzr: Ohetynel, Gvzr: 2022-03-12 14:30, Ybpngvba: 123 Ryz Fgerrg, Bssvpre Fzvgu bofreirq fhfcvpvbhf npgvivgl arne 5gu Nir ba 2022-03-13.}

        (2) \textbf{Atbash}: \textit{Hfhkvxg: Qlmzgszm, Xirnv: Yfitozib, Grnv: 2022-03-12 14:30, Olxzgrlm: 123 Von Hgivvg, Luurxvi Hnrgs lyhvievw hfhkrxrlfh zxgrergb mvzi 5gs Zev lm 2022-03-13.}

        (3) \textbf{Polybius}: \textit{41 43 41 34 15 13 42 :   24 33 32 11 42 22 11 32 ,   13 36 23 31 15 :   12 43 36 21 26 11 36 51 ,   
        42 23 31 15 :   2 0 2 2 - 0 3 - 1 2   1 4 : 3 0 ,   26 33 13 11 42 23 33 32 :   1 2 3   15 26 31   41 42 36 15 15 42 ,   
        33 16 16 23 13 15 36   41 31 23 42 22   33 12 41 15 36 44 15 14   41 43 41 34 23 13 23 33 43 41   11 13 42 23 44 23 42 51   
        32 15 11 36   5 42 22   11 44 15   33 32   2 0 2 2 - 0 3 - 1 3.}

        (4) \textbf{Vigenère}: \textit{Swdpgnt: Jqyavsap, Eciop: Mutrlccy, Tkxe: 2022-03-12 14:30, Lqnavtop: 123 Plo Svcege, Zfhtcgc Uxivs qmsgcvgo ufsrtckzuu aeeixtta nglr 5tj Axp qy 2022-03-13.}

        (5) \textbf{Reverse}: \textit{31-30-2202 no evA ht5 raen ytivitca suoicipsus devresbo htimS reciffO ,teertS mlE 321 :noitacoL ,03:41 21-30-2202 :emiT ,yralgruB :emirC ,nahtanoJ :tcepsuS.}

        (6) \textbf{SwapPairs}: \textit{uSpsce:tJ notaah,nC irem :uBgralyr ,iTem :02220--3211 :403 ,oLacitno :21 3lE mtSerte ,fOifec rmSti hboesvrdes suipicuo scaitivytn ae rt5 hvA eno2 20-2301-3.}

        (7) \textbf{DualAvgCode}: \textit{RTtvrtoqdfbdsu: IKnpmoaasugiaamo, BDqshjlndf: ACtvqsfhkmaaqsxz, SUhjlndf: 2022-03-12 14:30, KMnpbdaasuhjnpmo: 123 DFkmln RTsuqsdfdfsu, NPegeghjbddfqs RTlnhjsugi npacrtdfqsuwdfce rttvrtoqhjbdhjnptvrt aabdsuhjuwhjsuxz modfaaqs 5sugi AAuwdf npmo 2022-03-13.}

        (8) \textbf{ParityShift}: \textit{Rtrqdbu: Knozuizo, Bshld: Ctsfmzsx, Uhld: 2022-03-12 14:30, Mnbzuhno: 123 Dml Rusddu, Ngghbds Rlhui ncrdswde rtrqhbhntr zbuhwhux odzs 5ui Zwd no 2022-03-13.}

        (9) \textbf{WordShift}: \textit{pect:Sus athan,Jon me:Cri glary,Bur e:Tim 2-03-12202 30,14: ation:Loc 123 Elm eet,Str icerOff thSmi ervedobs picioussus ivityact rnea 5th Ave no 2-03-13202.}

    \end{exmp}
\end{figure*}

\begin{figure*}[p]
    \centering
    \begin{exmp}{Plain-Ciphertext Pair (Health Information) - Letter\&Number}{parallel}
        \small
        \setstretch{1.4}  

        \# Domain: Personal Privacy Data

        \#\# Subdomain: Health Information

        \#\#\# Tag Combination: ["Patient ID", "Diagnosis Records"]  

        \textbf{Plaintext}:
        
        \textit{Patient ID: R094713; Name: Jamie Lee; Age: 45; Gender: Female; EMR: EHR-234987.} 

        \textbf{Encryption results:}
        
        (1) \textbf{Rot13}: \textit{Cngvrag VQ: E327046; Anzr: Wnzvr Yrr; Ntr: 78; Traqre: Srznyr; RZE: RUE-567210.}

        (2) \textbf{Atbash}: \textit{Kzgrvmg RW: I905286; Mznv: Qznrv Ovv; Ztv: 54; Tvmwvi: Uvnzov; VNI: VSI-765012.}

        (3) \textbf{Polybius}: \textit{34 11 42 23 15 32 42   23 14 :   36 66 65 56 63 53 55 ;   
        32 11 31 15 :   24 11 31 23 15   26 15 15 ;   11 21 15 :   56 61 ;   
        21 15 32 14 15 36 :   16 15 31 11 26 15 ;   15 31 36 :   15 22 36 - 54 55 56 65 64 63.}

        (4) \textbf{Reverse}: \textit{.789432-R HRE ;elameF :redneG ;54 :egA ;eeL eimaJ :emaN ;317490 R :DI tneitaP}

        (5) \textbf{SwapPairs}: \textit{aPtetni DI: 0R94713; aNme: aJmei eLe; gAe: 45; eGndre: eFmale; MRE: HRE-239487.}

        (6) \textbf{WordShift}: \textit{atientP ID: R94713; ameN: Jamie eLe; geA: 45; enderG: emaleF; REM: EHR-234987.}

        (7) \textbf{DualAvgCode}: \textit{OQaasuhjdfmosu HJCE: QS009935680224; MOaalndf: IKaalnhjdf KMdfdf; AAfhdf: 3546; FHdfmocedfqs: EGdflnaakmdf; DFLNQS: DFGIQS-132435997968.}

        (8) \textbf{ParityShift}: \textit{Qzuhdou HE: S185602; Ozld: Kzlhd Mdd; Zfd: 54; Fdoeds: Gdlzmd; DLS: DIS-325896.}

    \end{exmp}
\end{figure*}

\begin{figure*}[p]
    \centering
    \begin{exmp}{Plain-Ciphertext Pair (Banking Information) - Letter\&Number}{parallel}
        \small
        \setstretch{1.4}  

        \# Domain: Financial Confidential Data

        \#\# Subdomain: Banking Information

        \#\#\# Tag Combination: ["Account Number", "Bank Card Number","Payment Platform ID"]  

        \textbf{Plaintext}:
        
        \textit{Account Number: 123456789, Bank: LA Bank, Card Number: 9876-5432-1098-7654, Payment Method: Virtual Credit Card, Payment Platform ID: ABC123XYZ, Timestamp: 2023-09-15 14:35, Amount: \$250.00.} 

        \textbf{Encryption results:}
        
        (1) \textbf{Rot13}: \textit{Nppbhag Ahzore: 456789012, Onax: YN Onax, Pneq Ahzore: 2109-8765-4321-0987, Cnlzrag Zrgubq: Iveghny Perqvg Pneq, Cnlzrag Cyngsbez VQ: NOP456KLM, Gvzrfgnzc: 5356-32-48 47:68, Nzbhag: \$583.33.}

        (2) \textbf{Atbash}: \textit{Zxxlfmg Mfnyvi: 876543210, Yzmp: OZ Yzmp, Xziw Mfnyvi: 0123-4567-8901-2345, Kzbnvmg Nvgslw: Erigfzo Xivwrg Xziw, Kzbnvmg Kozgulin RW: ZYX876CBA, Grnvhgznk: 7976-90-84 85:64, Znlfmg: \$749.99.}

        (3) \textbf{Polybius}: \textit{11 13 13 33 43 32 42   32 43 31 12 15 36 :   53 54 55 56 61 62 63 64 65 ,   
        12 11 32 25 :   26 11   12 11 32 25 ,   13 11 36 14   32 43 31 12 15 36 :   65 64 63 62 - 61 56 55 54 - 53 66 65 64 - 63 62 61 56 ,   
        34 11 51 31 15 32 42   31 15 42 22 33 14 :   44 23 36 42 43 11 26   13 36 15 14 23 42   13 11 36 14 ,   
        34 11 51 31 15 32 42   34 26 11 42 16 33 36 31   23 14 :   11 12 13 53 54 55 46 51 52 ,   
        42 23 31 15 41 42 11 31 34 :   54 66 54 55 - 66 65 - 53 61   53 56 : 55 61 ,   
        11 31 33 43 32 42 :   \$ 54 61 66 . 66 66 .}

        (4) \textbf{Vigenère}: \textit{Swdpgnt: Jqyavsap, Eciop: Mutrlccy, Tkxe: 2022-03-12 14:30, Lqnavtop: 123 Plo Svcege, Zfhtcgc Uxivs qmsgcvgo ufsrtckzuu aeeixtta nglr 5tj Axp qy 2022-03-13.}

        (5) \textbf{Reverse}: \textit{.00.052\$ :tnuomA ,53:41 51-90-3202 :pmatsemit ,ZYX321CBA :DI mroftalP tnemyap ,draC tiderC lautriV :dohtem tnemyap ,4567-8901-2345-6789 :rebmuN draC ,knaB AL :knaB ,987654321 :rebmuN tnuoccA}

        (6) \textbf{SwapPairs}: \textit{cAotcnu mNuber: 214365879, aBnk: A Lank, aCrd Nmu:bre 8967-5423-1980-7564, aPymnet Mtohed: Vritaul Cerdti aCdr, aPymnet Ptaforml DI: BAC321YXZ, iTmsetamp: 3202-90-51 53:41, aAmount: \$250.00.}

        (7) \textbf{DualAvgCode}: \textit{AAbdbdnptvmosu MOtvlnacdfqs: 021324354657687999, ACaamojl: KMAA ACaamojl, BDaaqsce MOtvlnacdfqs: 99796857-46352413-02009979-68574635, OQaaxzlndfmosu LNdfsuginpce: UWhjqssutvaakm BDqsdfcehjsu BDaaqsce, OQaaxzlndfmosu OQkmaasuegnpqsln HJCE: AAACBD021324WYXZZZ, SUhjlndfrtsuaalnoq: 13001324-0099-0246 0235:2446, AAlnnptvmosu: \$134600.0000.}

        (8) \textbf{ParityShift}: \textit{Zbbntou Otlcds: 032547698, Czoj: MZ Czoj, Bzse Otlcds: 8967-4523-0189-6745, Qzxldou Lduine: Whsutzm Bsdehu Bzse, Qzxldou Qmzugnsl HE: ZCB032YXA, Uhldruzlq: 3132-18-04 05:24, Zlntou: \$341.11.}
    \end{exmp}
\end{figure*}

\subsection{Detailed Descriptions of Encryption Algorithms}

\begin{table*}[p]
    \centering
    \renewcommand{\arraystretch}{1.3} 
    \caption{Descriptions of Encryption Algorithms in CipherBank}
    \label{tab:encryption_algorithms}
    \resizebox{\textwidth}{!}{%
    \begin{tabular}{l p{15cm}}
        \toprule
        \textbf{Algorithm} & \textbf{Description} \\
        \midrule
        \textbf{Rot13} & A simple \textit{substitution cipher} that shifts each letter 13 places forward in the alphabet. Encryption and decryption are identical, as applying the transformation twice restores the original text. Non-alphabetic characters remain unchanged. 
        
        Additionally, Rot13 in CipherBank supports number encryption by shifting digits cyclically within the range 0-9. \\
        \midrule
        \textbf{Atbash} & A \textit{monoalphabetic substitution cipher} where each letter is replaced with its counterpart from the reversed alphabet (e.g., A→Z, B→Y). Since the transformation is symmetric, encryption and decryption follow the same process. 
        
        CipherBank's Atbash implementation extends this to digits, where each number is replaced with its complement relative to 9 (e.g., 0→9, 1→8, ..., 9→0). \\
        \midrule
        \textbf{Polybius} &A \textit{fractionating substitution cipher} that replaces each letter with a two-digit coordinate from a 6×6 grid, mapping characters to numerical positions. Traditional Polybius squares typically use a 5×5 grid, supporting only letter encryption while merging I and J into the same cell, leading to ambiguity during decryption. To address this limitation and enable number encryption, CipherBank extends the Polybius square to a 6×6 grid, allowing both letters and numbers to be uniquely represented as coordinate pairs, increasing the cipher's complexity. By default, the key is set to \texttt{"ACL"}.\\
        \midrule
        \textbf{Vigenère} & A \textit{polyalphabetic substitution cipher} that employs multiple shifting alphabets determined by a repeating key. Unlike monoalphabetic ciphers that use a single mapping, Vigenère utilizes multiple substitution tables, where each plaintext letter is shifted based on the corresponding key character’s position in the alphabet. 
        
        This multi-table approach enhances security by distributing letter frequencies across different shifts, making it more resistant to frequency analysis. Decryption reverses this process by applying the inverse shifts dictated by the key. Unlike Rot13, it requires a key for both encryption and decryption. \\
        \midrule
        \textbf{Reverse} & A \textit{transposition cipher} that reverses the order of all characters in the plaintext. Since it does not substitute characters, it preserves all information but alters the sequence, making it effective against naive attacks. \\
        \midrule
        \textbf{SwapPairs} & A \textit{transposition cipher} that swaps adjacent characters in the plaintext. If the text length is odd, the final character remains unchanged. Decryption follows the same swapping process. \\
        \midrule
        \textbf{DualAvgCode} & A \textit{custom transformation} where each letter expands into two adjacent characters, shifting one position forward and one position backward in the ASCII table. Special cases (e.g., 'a', 'z', 'A', 'Z') are duplicated instead. 
        
        CipherBank extends this method to digits, where each number expands into two adjacent values (e.g., 2 → "13", 5 → "46"), increasing redundancy in the encrypted text. \\
        \midrule
        \textbf{ParityShift} & A \textit{custom encryption method} that shifts each letter one position forward or backward based on its ASCII parity. Even-ASCII characters shift forward, while odd-ASCII characters shift backward.
        
        For digits, ParityShift follows a similar rule, shifting numbers based on their parity (e.g., even numbers shift up, odd numbers shift down within 0-9). \\
        \midrule
        \textbf{WordShift} & A transformation applied at the word level rather than the character level. Each word undergoes a left shift by a fixed number of positions, cycling characters within the word while preserving word spacing. Decryption reverses this shift, ensuring character order is restored within each word. By default, the shift is set to 3 positions.\\
        \bottomrule
    \end{tabular}
    }
\end{table*}

This section provides detailed descriptions of the nine encryption algorithms used in CipherBank. These algorithms span substitution, transposition, and custom-designed ciphers, covering a range of complexity levels. Notably, Rot13, Atbash, Polybius, DualAvgCode, and ParityShift also support numeric encryption, further enhancing the diversity of decryption challenges. \textbf{Table \ref{tab:encryption_algorithms}} outlines each algorithm and its transformation rules.Some detailed encryption examples are provided below, illustrating how different ciphers transform plaintext into ciphertext.

For each encryption algorithm, we have implemented a corresponding decryption algorithm to ensure that ciphertext can be fully restored to its original plaintext. This guarantees the reversibility and integrity of the encryption schemes used in CipherBank, allowing for a rigorous evaluation of model decryption capabilities. The decryption process follows the exact inverse of the encryption transformations, ensuring consistency across all test cases.

\section{Experimental Setup Details}
In our evaluation, we adopt a 3-shot approach. A more natural Ciphertext-Only Attack (zero-shot) setting was not adopted, as it would reduce the task to brute-force decryption, where the model blindly applies all known encryption algorithms in search of a coherent output. This contradicts the goal of reasoning-based inference, where the model is expected to deduce encryption rules from provided examples rather than rely on exhaustive trial and error.  

To ensure a balanced evaluation of decryption difficulty, substitution ciphers exclude numbers to prevent inconsistencies arising from differing cyclic structures. In contrast, ciphers that do not involve direct substitution, such as Reverse, WordShift, and similar methods, process numbers normally, preserving structural integrity within the encrypted text.

For all open-source models, we conduct evaluations using the OpenCompass\footnote{\url{https://github.com/open-compass/opencompass}} framework with default temperature to ensure consistent outputs. For models evaluated via API, \textbf{we perform $5$ independent test runs} per model and report the average result to enhance stability and reliability.

\subsection{Prompts Used for Querying}
\label{sec:prompt}

\begin{figure*}[p]
    \centering
    \begin{exmp}{}{parallel}
        \small
        \setstretch{1.4}  

        \textbf{\#\# Role:}

        \hspace{2em}Cryptography Analysis Expert.

        \textbf{\#\# Goals:}

        \hspace{2em}Utilize the provided ciphertext and plaintext examples to analyze encryption patterns and decrypt new ciphertext.

        \textbf{\#\# Workflow:}

        \hspace{2em}1. Analyze the provided ciphertext and plaintext examples to identify possible encryption patterns and rules.

        \hspace{2em}2. Apply the decryption algorithm to the new ciphertext, attempt to decrypt, and verify the results.

    \end{exmp}
    \caption{System Prompt}
    \label{tab:system_prompt}
\end{figure*}

\begin{figure*}[p]
    \centering
    \begin{exmp}{}{parallel}
        \small
        \setstretch{1.4}  

        \textbf{\#\# Background:}

        \hspace{2em}Decrypt the provided Ciphertext and return the corresponding Plaintext following the given examples.

        \hspace{2em}Think step by step.

        \hspace{2em}Provide the Plaintext result in the format <result>text</result>, where text is the decrypted text.

        \textbf{\#\# Examples:}

        \hspace{2em}* Example 1:
        
            \hspace{4em}- Ciphertext:
            
                \hspace{6em}Anzr: Rzvyl Wbuafba; Qngr bs Ovegu: Whyl 15, 1990; Cnffcbeg Ahzore: L987654321
                
            \hspace{4em}- Plaintext:
            
                \hspace{6em}Name: Emily Johnson; Date of Birth: July 15, 1990; Passport Number: Y987654321
                
        \hspace{2em}* Example 2:
        
            \hspace{4em}- Ciphertext:
            
                \hspace{6em}Pnfr Ahzore: 2024-CF-001234; Pnfr Glcr: Gursg/Oernx-Va; Svyvat Qngr: Bpgbore 19, 2024
            
            \hspace{4em}- Plaintext:
            
                \hspace{6em}Case Number: 2024-PS-001234; Case Type: Theft/Break-In; Filing Date: October 19, 2024
                
        \hspace{2em}* Example 3:
        
            \hspace{4em}- Ciphertext:
            
                \hspace{6em}gnk$\_$vqragvsvre: GKA-2023-NOP456, gnk\_erpbeqf: {lrne: 2023, fgnghf: Cebprffrq, ershaq\_vffhrq: 620.00}
            
            \hspace{4em}- Plaintext:
            
                \hspace{6em}tax$\_$identifier: TXN-2023-ABC456, tax\_records: {year: 2023, status: Processed, refund\_issued: 620.00}

        \textbf{\#\# Input:}

        \hspace{2em}- Ciphertext:
        
            \hspace{4em}Yrqvn, na Nzrevpna, erfvqrf va Ybf Natryrf.
            
        \hspace{2em}- Plaintext:
        
    \end{exmp}
    \caption{User Prompt (Rot13 - 3shot - Only Letter)}
    \label{tab:user_prompt_1}
\end{figure*}

This section outlines the prompts used to query models during evaluation. To ensure consistency, all models were tested under a 3-shot setting, where they were provided with three plaintext-ciphertext pairs before attempting to decrypt a new ciphertext. The prompts were designed to encourage logical inference rather than relying on prior knowledge, guiding models to extract encryption patterns and apply the learned rules systematically. Below, \textbf{Figure~\ref{tab:system_prompt} }provides the system prompt (some reasoning models may not support system prompts), while\textbf{ Figure~\ref{tab:user_prompt_1}} present the detailed user prompts.

\subsection{Post-processing Methods}

During querying, we instruct the model to think step by step and enclose the final decrypted output within \textbf{<result>...</result>} tags. To extract the decoded plaintext, we apply the regular expression \textbf{'<result>(.*?)</result>'}, capturing the content between these tags. The matching process is \textbf{case-insensitive}, aligning with algorithms like Polybius, which inherently do not differentiate between uppercase and lowercase letters when restoring plaintext. This ensures consistency across different decryption schemes.

\section{Extended Experimental Results}
\subsection{Levenshtein Distance Evaluation from Main Results}
\label{LD_result}
In the main text, most reported results are based on \textit{accuracy}, which provides a binary assessment of decryption success. However, accuracy does not account for cases where decrypted outputs closely resemble the ground truth but contain minor errors. To provide a more fine-grained evaluation, we also compute \textit{Levenshtein similarity}, which measures the edit distance between the model output and the correct plaintext.

We define the \textbf{Levenshtein similarity score} as follows:

\begin{equation}
S_{\text{lev}} = 1 - \frac{d_{\text{lev}}(P_{\text{pred}}, P_{\text{ref}})}{\max(|P_{\text{pred}}|, |P_{\text{ref}}|)}
\end{equation}

where:
\begin{itemize}
    \item $d_{\text{lev}}(P_{\text{pred}}, P_{\text{ref}})$ is the \textit{Levenshtein distance} between the predicted and reference plaintexts.
    \item $|P_{\text{pred}}|$ and $|P_{\text{ref}}|$ denote the lengths of the predicted and reference plaintexts, respectively.
\end{itemize}

This metric \textit{normalizes the edit distance} by the length of the longer string, ensuring that similarity is measured on a scale from $0$ to $1$, where $1$ represents an exact match and lower values indicate increasing deviations from the ground truth.

\begin{table*}[p]
    \centering
    \caption{Results on CipherBank(3-shot) Levenshtein similarity}
    \label{tablel}
    \resizebox{\textwidth}{!}
    {
    \begin{tabular}{lcccccccccccc}
        \toprule
        \multirow{2}{*}{\textbf{Model}} & \multicolumn{4}{c}{\textbf{Substitution Ciphers}} & \multicolumn{2}{c}{\textbf{Transposition Ciphers}}  & \multicolumn{3}{c}{\textbf{Custom Ciphers}}  & \multicolumn{1}{c}{\textbf{Cipher Score}}  \\
        \cmidrule(lr){2-5} \cmidrule(lr){6-7} \cmidrule(lr){8-10} \cmidrule(lr){11-11} 
        &Rot13 & At ba sh & Polybius & Vigenère & Reverse & SwapPairs  & DualAvgCode & ParityShift & WordShift & Levenshtein Similarity \\
        \midrule
        \rowcolor[HTML]{f8e2d1} 
        \multicolumn{11}{c}{\textit{Open-source Chat Models}} \\ \midrule
        Mixtral-8x22B-v0.1   & 0.4542 & 0.3744 & 0.2694 & 0.4032 & 0.3810 & 0.4745 & 0.3330 & 0.3871 & 0.6401 & 0.4130\\
        Qwen2.5-72B-Instruct & 0.5556 & 0.4288 & 0.3042 & 0.4016 & 0.4022 & 0.5308 & 0.3718 & 0.4786 & 0.8427 & 0.4796\\ 
        Llama-3.1-70B-Instruct  & 0.5776 & 0.4378 & 0.3132 & 0.4431 & 0.3775 & 0.5542 & 0.3990 & 0.4505 & 0.7288 & 0.4758 \\ 
        Llama-3.3-70B-Instruct & 0.5754 & 0.4054 & 0.1317 & 0.4397 & 0.2482 & 0.5375 & 0.3833 & 0.4096 & 0.4580 & 0.3988\\ 
        DeepSeek-V3 &  0.9195 & 0.7594 & 0.4562 & 0.4844 & \underline{0.9088} & 0.6975 & 0.4205 & 0.5731 & 0.8887 & 0.6787 \\  \midrule
        \rowcolor[HTML]{fdf3d0}
        \multicolumn{11}{c}{\textit{Closed-source Models}} \\ \midrule
        GPT-4o-mini-2024-07-18 & 0.6459 & 0.4935 & 0.2463 & 0.4499 & 0.5664 & 0.6005 & 0.3418 & 0.4188 & 0.7258 & 0.4988\\
        GPT-4o-2024-08-06 & 0.9603 & 0.5876 & 0.3445 & 0.5346 & 0.8170 & 0.7968 & 0.4304 & 0.5850 & 0.8940 & 0.6612\\
        GPT-4o-2024-11-20 & 0.9340 & 0.6054 & 0.3511 & 0.5338 & 0.7277 & 0.6780 & 0.4235 & 0.5530 & 0.8715 & 0.6309\\
        gemini-1.5-pro &  0.9309 & 0.5043 & 0.4969 & 0.5201 & 0.7536 & 0.7317 & 0.4784 & 0.5720 & 0.8819 & 0.6522 \\
        gemini-2.0-flash-exp & 0.9616 & 0.6567 & 0.4813 & 0.5064 & 0.8901 & 0.7569 & 0.4476 & 0.5308 & 0.8605 & 0.6769 \\
        Claude-Sonnet-3.5-1022 & \textbf{0.9984} & \textbf{0.9961 }& \textbf{0.9955} & \textbf{0.7143} & \textbf{0.9893} & \textbf{0.9262} & 0.7874 & \textbf{0.9883 }&\textbf{ 0.9712 }& \textbf{0.9296} \\
        \midrule
        \rowcolor[HTML]{d6ecda} 
        \multicolumn{11}{c}{\textit{ Reasoning Models}} \\ \midrule
        QwQ-32B-Preview  & 0.2477 & 0.1591 & 0.1231 & 0.1660 & 0.1444 & 0.1666 & 0.1564 & 0.1645 & 0.3057 & 0.1815\\
        DeepSeek-R1  & 0.9920 & 0.9761 & 0.9344 & 0.5227 & 0.7368 & 0.7213 & \underline{0.8316} & 0.6928 & 0.8491 & 0.8063\\ 
        gemini-2.0-flash-thinking  & 0.9664 & 0.8571 & 0.9074 & 0.5511 & 0.8508 & 0.7788 & 0.4261 & \underline{0.7353} & 0.8777 & 0.7723 \\
        o1-mini-2024-09-12 & \underline{0.9757} & 0.9860 & 0.9563 & 0.5412 & 0.5959 & 0.5267 & 0.3954 & 0.6935 & 0.7236 & 0.7105\\
        o1-2024-12-17  & 0.8320 & \underline{0.9928} & \underline{0.9640} & \underline{0.5642} & 0.7725 & \underline{0.9208} & \textbf{0.8653} & 0.6562 & \underline{0.9335} & \underline{0.8335}\\
        \midrule
        \bottomrule
    \end{tabular}%
    }
\end{table*}

\begin{figure*}[p] 
    \centering
    \includegraphics[width=0.92\textwidth]{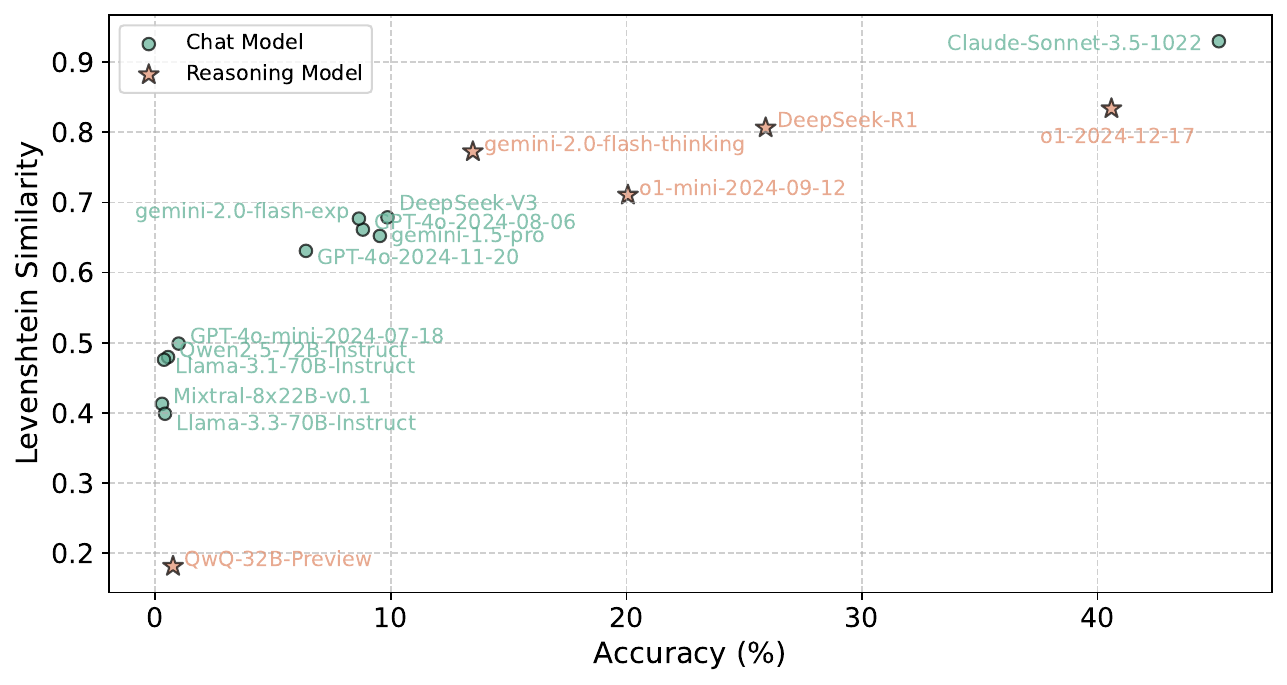}  
    \caption{Model Performance - Accuracy vs. Levenshtein Similarity.}
    \label{fig:xy}
\end{figure*}

The corresponding \textbf{Levenshtein-based evaluation results} for \textbf{Table~\ref{table2}} are presented in \textbf{Table~\ref{tablel}} and \textbf{Figure~\ref{fig:xy}}, offering deeper insights into models' decryption performance beyond strict accuracy metrics.

One key observation is that most models achieve significantly higher Levenshtein similarity scores than their accuracy scores, indicating that even when decryption is incorrect, outputs often retain structural similarities to the original plaintext. This suggests that models capture some encryption patterns but struggle with full decryption, failing to consistently apply correct transformations. Notably, Claude-Sonnet-3.5 achieves near-perfect scores (>0.99 for most ciphers), demonstrating its ability to minimize decryption errors while maintaining structural accuracy, making it the most reliable model overall.  

Interestingly, reasoning models such as DeepSeek-R1 and o1 exhibit a large gap between accuracy and Levenshtein similarity. Despite their moderate accuracy, their similarity scores often exceed 0.80, indicating that they frequently produce outputs that preserve much of the original structure but contain systematic errors. This suggests that reasoning models are better at capturing encryption logic but may struggle with precise execution, sometimes overcomplicating simpler tasks.  

Conversely, chat models such as DeepSeek-V3 and Llama-based models exhibit high variability, showing relatively low accuracy but moderate Levenshtein similarity (0.40 - 0.70). This indicates a tendency toward semantic approximation rather than strict decryption, where models generate linguistically plausible outputs that fail to adhere to precise encryption rules.  

Another notable trend is that transposition ciphers (e.g., Reverse, SwapPairs) yield lower Levenshtein similarity scores across all models, confirming that character reordering remains a major challenge. Unlike substitution ciphers, where models can rely on token-level mappings, transposition ciphers require strict positional tracking, which even the strongest models struggle to handle effectively.  

Overall, Levenshtein similarity results highlight fundamental differences in how chat and reasoning models approach decryption. Chat models rely more on semantic fluency, leading to structurally incorrect but coherent outputs, whereas reasoning models exhibit stronger pattern retention but occasionally fail due to overgeneralization or overthinking. These findings suggest that while LLMs can approximate decryption rules, achieving precise symbolic transformations remains a significant challenge, especially for positional-based ciphers.

\subsection{Additional Analysis and Insights}
\label{appendix_analysis}
In this section, we present more detailed experimental results that complement the findings in the main text. These additional analyses provide further insights into model performance across different encryption schemes, highlighting trends, challenges, and specific cases where models excel or struggle.

\begin{table*}[p]
    \centering
    \caption{Decryption Performance on Short Texts}
    \label{table3_short}
    \resizebox{\textwidth}{!}
    {
    \begin{tabular}{lcccccccccccc}
        \toprule
        \multirow{2}{*}{\textbf{Model}} & \multicolumn{4}{c}{\textbf{Substitution Ciphers}} & \multicolumn{2}{c}{\textbf{Transposition Ciphers}}  & \multicolumn{3}{c}{\textbf{Custom Ciphers}}  & \multicolumn{1}{c}{\textbf{Cipher Score}}  \\
        \cmidrule(lr){2-5} \cmidrule(lr){6-7} \cmidrule(lr){8-10} \cmidrule(lr){11-11} 
        &Rot13 & Atbash & Polybius & Vigenère & Reverse & SwapPairs  & DualAvgCode & ParityShift & WordShift & Accuracy$_{\text{avg}}$\\
        \midrule
        DeepSeek-V3 & 40.00 &  27.83 & 4.35 & 1.74 & 29.57 & 0.87  & 0.87 & 2.61 & 11.3 & 13.24  \\ 
        DeepSeek-R1 & 80.00 &  71.30 & 53.04 & 0.87 & 18.26 & 0.87  & 35.65 & 18.26 & 12.17 & 32.27  \\ 
        GPT-4o-2024-11-20 & 34.78 &  13.04 & 0.87 & 0 & 21.74 & 1.74  & 0.87 & 1.74 & 10.43 & 9.47 \\
        gemini-2.0-flash-exp & 42.61 &  4.35 & 1.74 & 0.87 & 40.87 & 2.61  & 0 & 1.74 & 8.70 & 11.50\\
        Claude-Sonnet-3.5-1022 &86.09 &  77.39 & 69.57 & 3.48 & 77.39 & 8.70  & 9.57 & 63.48 & 42.61 & 48.70  \\
        gemini-2.0-flash-thinking & 52.17 &  26.96 & 33.91 & 2.61 & 33.91 & 0.87  & 0 & 13.91 & 14.78 & 19.90\\
        o1-mini-2024-09-12 & 64.35 &  82.61 & 65.22 & 0 & 15.65 & 0  & 6.67 & 13.91 & 2.61 & 33.77 \\
        o1-2024-12-17  & 61.74 &  89.57 & 84.55 & 0.87 & 23.48 & 46.67  & 61.74 & 17.17 & 35.80 &47.61 \\ 
        \midrule
        \bottomrule
    \end{tabular}%
    }
\end{table*}

\begin{table*}[p]
    \centering
    \caption{Decryption Performance on Long Texts}
    \label{table3_long}
    \resizebox{\textwidth}{!}
    {
    \begin{tabular}{lcccccccccccc}
        \toprule
        \multirow{2}{*}{\textbf{Model}} & \multicolumn{4}{c}{\textbf{Substitution Ciphers}} & \multicolumn{2}{c}{\textbf{Transposition Ciphers}}  & \multicolumn{3}{c}{\textbf{Custom Ciphers}}  & \multicolumn{1}{c}{\textbf{Cipher Score}}  \\
        \cmidrule(lr){2-5} \cmidrule(lr){6-7} \cmidrule(lr){8-10} \cmidrule(lr){11-11} 
        &Rot13 & Atbash & Polybius & Vigenère & Reverse & SwapPairs  & DualAvgCode & ParityShift & WordShift & Accuracy$_{\text{avg}}$\\
        \midrule
        DeepSeek-V3 & 26.53 &  4.76 & 0.68 & 0 & 9.52 & 0  & 0 & 0 & 5.44 & 5.22  \\ 
        DeepSeek-R1 & 68.03 &  48.98 & 37.41 & 0 & 4.76 & 0  & 14.97 & 8.84 & 5.44 & 20.94  \\ 
        GPT-4o-2024-11-20 & 20.41 &  4.08 & 0 & 0 & 12.24 & 0  & 0 & 0 & 3.40 & 4.46 \\
        gemini-2.0-flash-exp & 30.61 &  2.04 & 1.36 & 0 & 20.41 & 0.68  & 0 & 0 & 2.72 & 6.42\\
        Claude-Sonnet-3.5-1022 &92.52 &  78.91 & 82.31 & 1.36 & 63.95 & 5.44  & 2.72 & 63.27 & 40.14 & 47.85  \\
        gemini-2.0-flash-thinking & 31.29 &  9.52 & 12.24 & 0 & 14.29 & 1.36  & 0 & 2.72 & 4.76 & 8.47\\
        o1-mini-2024-09-12 & 31.97 &  57.14 & 32.65 & 0 & 0 & 0  & 0 & 2.72 & 0 & 17.35 \\
        o1-2024-12-17  & 58.50 &  70.75 & 61.11 & 0.68 & 8.16 & 15.38  & 41.5 & 8.66 & 25.66 & 34.38 \\ 
        \midrule
        \bottomrule
    \end{tabular}%
    }
\end{table*}

\begin{figure*}[p]
    \centering
    \begin{exmp}{Plaintext Examples}{parallel}
        \small
        \setstretch{1.4}  
        
        \textbf{Short:} James, American, is married to Susan.  

        \textbf{Long:} John Smith, born on January 15, 1990, holds American nationality and resides at 123 Elm Street, Springfield, Illinois. His mobile number is +1-312-555-6789, and his ID card number is IDURITY1234567. He is married to Jane Smith, who can be reached at +1-312-555-6789. They have two children: Emily (16, high school) and Michael (12, middle school). Their address and contact information are the same.  

        \vspace{2mm} 
        
        \textbf{Short:} Jimmy, GPA: 3.71. 

        \textbf{Long:} David Wilson, Masters in Data Science, GPA: 3.95, Expected Graduation: 2023, Courses: Big Data Analytics, Machine Learning, Data Visualization.
        
        \vspace{2mm}
        \textbf{Short:} Medical Record Number: 987-654-321; Patient Name: James. 

        \textbf{Long:} David Wilson, Masters in Data Science, GPA: 3.95, Expected Graduation: 2023, Courses: Big Data Analytics, Machine Learning, Data Visualization.

        \vspace{2mm}
        \textbf{Short:} Lucas, lucas@ucc.company.com

        \textbf{Long:} Hank, Senior Developer, IT Department, Salary: \$95,000, Bonuses: \$5,000, Allowances: \$2,000 (Remote Work), Performance Rating: A, Full-time, Start Date: 2020-03-15, Last Promotion: 2021-08-10, Benefits: Health Insurance, Retirement 5\%, Training: \$1,500/year, Projects: Nexus, Zeta, Feedback: 4.5/5        
    \end{exmp}
    \caption{Samples used for length sensitivity analysis}
    \label{length_example}
\end{figure*}

In the analysis of length sensitivity, plaintexts of different lengths can be seen in \textbf{Figure~\ref{length_example}.} The impact of plaintext length on decryption performance is shown in \textbf{Table~\ref{table3_short} }and \textbf{Table~\ref{table3_long}}, where we compare model accuracy on short vs. long texts. These results illustrate how increasing text length affects model performance, revealing notable differences in decryption robustness across various architectures

\begin{figure*}[h!]
    \centering
    \begin{exmp}{Noise Example}{parallel}
        \small
        \setstretch{1.4}  

        \textbf{Example 1:}

        \textbf{Origin:} Card Number: 9876 5432 1098 7654

        \textbf{Noise:} Card Numbr: 9876 54-32 1O98 765four  

        \vspace{2mm} 

        \textbf{Example 2:}

        \textbf{Origin:} Pay Date: 2023-05-15, Income: \$75,000, Currency: USD, Bonus: \$5,000

        \textbf{Noise:} Pay Date (scheduled): 2023-05-15! Income approx: \$75,000. Currency spec: USD, and Bonus ~= \$5,000.

        \vspace{2mm} 

        \textbf{Example 3:}

        \textbf{Predictions:} Officer ID: P12345, Name: John, Position: Sergeant, Department: Homicide  

        \textbf{References:} Officer Identification-No.: P12345, Full-Name: John (J.), Job-Title: Sergeant, Dept.: Homicide Squad.

    \end{exmp}
    \caption{The samples used for the noise comparison experiments.}
     \label{noise_example}
\end{figure*}

\begin{table*}[h!]
    \centering
    \caption{Decryption Performance without Noise}
    \label{table_noise_short}
    \resizebox{\textwidth}{!}{%
    \begin{tabular}{lcccccccc}
        \toprule
        \textbf{Model}  & Rot13 & Atbash & Reverse & SwapPairs  & ParityShift & WordShift & Accuracy$_{\text{avg}}$\\
        \midrule
        \multicolumn{8}{c}{\textit{Open-source Models}} \\ \midrule
        DeepSeek-V3 & 50.00 & 31.50 & 18.50 & 6.50 & 9.00 & 17.00 & 22.08 \\ 
        DeepSeek-R1 & 83.50 & 77.50 & 42.00 & 2.50 & 20.00 & 5.50 & 38.50 \\ \midrule
        \multicolumn{8}{c}{\textit{Closed-source Models}} \\ \midrule
        GPT-4o-2024-11-20 & 49.50 & 10.50 & 13.50 & 0 & 3.50 & 5.50 & 13.75 \\
        Gemini-2.0-flash-exp & 45.00 & 7.50 & 42.50 & 2.50 & 5.00 & 15.50 & 19.67 \\
        Claude-Sonnet-3.5-1022 & 92.50 & 85.00& 62.50 & 10.00 & 70.00 & 35.00 & 59.17 \\
        Gemini-2.0-flash-thinking & 62.50 & 33.50 & 22.50 & 0 & 17.50 & 1.50 & 22.92 \\
        o1-mini-2024-09-12 & 55.50 & 67.50 & 5.00 & 0 & 17.50 & 0 & 24.25 \\
        \midrule
        \bottomrule
    \end{tabular}%
    }
\end{table*}

\begin{table*}[h!]
    \centering
    \caption{Decryption Performance with Noise}
    \label{table_noise_long}
    \resizebox{\textwidth}{!}{%
    \begin{tabular}{lcccccccc}
        \toprule
        \textbf{Model}  & Rot13 & Atbash & Reverse & SwapPairs  & ParityShift & WordShift  & Accuracy$_{\text{avg}}$\\
        \midrule
        \multicolumn{8}{c}{\textit{Open-source Models}} \\ \midrule
        DeepSeek-V3 & 8.50 & 10.50 & 7.50 & 0 & 0.50 & 1.50 & 4.75 \\ 
        DeepSeek-R1 & 33.50 & 23.00 & 4.50 & 0 & 1.50 & 0 & 10.42 \\ \midrule
        \multicolumn{8}{c}{\textit{Closed-source Models}} \\ \midrule
        GPT-4o-2024-11-20 & 5.50 & 0 & 4.50 & 0 & 0 & 0 & 1.67 \\
        Gemini-2.0-flash-exp & 2.50 & 0 & 0 & 2.50 & 0 & 0 & 0.83 \\
        Claude-Sonnet-3.5-1022 & 50.50 & 40.00 & 20.00 & 2.50 & 30.00 & 7.50 & 25.08 \\
        Gemini-2.0-flash-thinking & 30.50 & 19.00 & 3.50 & 0 & 2.50 & 0 & 9.25 \\
        o1-mini-2024-09-12 & 15.00 & 20.0 & 0 & 0 & 0 & 0 & 5.83 \\
        \midrule
        \bottomrule
    \end{tabular}%
    }
\end{table*}

The dataset used for the noise interference experiments can be found in \textbf{Figure \ref{noise_example}}. Detailed results on the impact of noise on decryption performance are presented in \textbf{Table~\ref{table_noise_short}} and\textbf{ Table~\ref{table_noise_long}}, comparing model performance on short and long plaintexts under noisy conditions. These findings highlight the varying degrees of resilience across models, with some maintaining reasonable performance under noise while others degrade significantly.

\begin{figure*}[h!]
    \centering
    \begin{exmp}{}{parallel}
        \small
        \setstretch{1.4}  

        \textbf{\#\# Background:}

        \hspace{2em}Decrypt the provided Ciphertext and return the corresponding Plaintext following the given examples.

        \hspace{2em}Think step by step.

        \hspace{2em}Provide the Plaintext result in the format <result>text</result>, where text is the decrypted text.

        \textbf{\#\# Examples:}

        \hspace{2em}* Example 1:
        
            \hspace{4em}- Ciphertext:
            
                \hspace{6em}Mznv: Vnrob Qlsmhlm; Wzgv lu Yrigs: Qfob 84, 8009; Kzhhklig Mfnyvi: B012345678
                
            \hspace{4em}- Plaintext:
            
                \hspace{6em}Name: Emily Johnson; Date of Birth: July 15, 1990; Passport Number: Y987654321

        \hspace{2em}* Example 2:
        
            \hspace{4em}- Ciphertext:
            
                \hspace{6em}Xzhv Mfnyvi: 7975-KH-998765; Xzhv Gbkv: Gsvug/Yivzp-Rm; Urormt Wzgv: Lxglyvi 80, 7975
            
            \hspace{4em}- Plaintext:
            
                \hspace{6em}Case Number: 2024-PS-001234; Case Type: Theft/Break-In; Filing Date: October 19, 2024
                
        \hspace{2em}* Example 3:
        
            \hspace{4em}- Ciphertext:
            
                \hspace{6em}gzc\_rwvmgrurvi: GCM-7976-ZYX543, gzc\_ivxliwh: {bvzi: 7976, hgzgfh: Kilxvhhvw, ivufmw\_rhhfvw: 379.99}
            
            \hspace{4em}- Plaintext:
            
                \hspace{6em}tax\_identifier: TXN-2023-ABC456, tax\_records: {year: 2023, status: Processed, refund\_issued: 620.00}

        \textbf{\#\# Input:}

        \hspace{2em}- Ciphertext:
        
            \hspace{4em}Wvzm slowh gsv kzhhklig mfnyvi Z87654321.
            
        \hspace{2em}- Plaintext:
        
    \end{exmp}
    \caption{User Prompt (Atbash - 3shot - Letter \& Number)}
    \label{tab:user_prompt_3}

\end{figure*}

\begin{table*}[h!]
    \centering
    \caption{Impact of Encryption Scope on Decryption Performance}
    \label{table_number}
    \resizebox{\textwidth}{!}{%
    \begin{tabular}{lccccccc}
        \toprule
        \textbf{Model}  & Rot13 & Atbash & Polybius  & DualAvgCode & ParityShift  & Accuracy$_{\text{avg}}$\\
        \midrule
        \multicolumn{7}{c}{\textit{Open-source Models}} \\ \midrule
        DeepSeek-V3 &  68.94/23.32 & 24.02/14.64 & 19.35/6.01 & 3.51/0 & 11.31/0 & 25.23 / 8.79 \\ 
        DeepSeek-R1 &  59.10/43.05 & 63.19/23.02 & 39.21/43.23 & 37.36/0 & 13.05/0.76 & 42.38 / 22.01 \\ \midrule
        \multicolumn{7}{c}{\textit{Closed-source Models}} \\ \midrule
        GPT-4o-2024-11-20 & 27.53/0 & 10.08/0 & 0/0 & 2.54/0 & 2.67/0 & 8.56 / 0 \\
        gemini-2.0-flash-exp & 47.54/0 & 7.50/2.50 & 7.50/5.05 & 0/0 & 2.67/0 & 13.04 / 1.51 \\
        Claude-Sonnet-3.5-1022 & 92.50/50.00 & 87.56/27.53 & 65.00/32.25 & 15.00/0 & 62.54/17.35 & 64.52 / 25.43 \\
        gemini-2.0-flash-thinking & 35.00/2.65 & 0/2.54 & 0/10.00 & 0/0 & 2.50/0 & 7.50 / 3.04 \\
        o1-mini-2024-09-12 & 50.00/32.59 & 72.57/35.00 & 40.00/42.53 & 0/0 & 7.50/0.76 & 34.01 / 22.18 \\
        \midrule
        \bottomrule
    \end{tabular}%
    }
    
    \vspace{2mm} 
    \caption*{\small Note: Values before the `/` indicate performance when encrypting \textbf{letters only}, while values after the `/` represent performance when encrypting \textbf{both letters and numbers}.}
\end{table*}

In the analysis of the impact of encryption scope on decryption performance, the test prompts used are shown in \textbf{Figure~\ref{tab:user_prompt_3}}. Detailed results are presented in \textbf{Table~\ref{table_number}}. This analysis compares model performance when encrypting only letters versus encrypting both letters and numbers. The results highlight how different models handle the increased complexity introduced by number encryption, showing varying degrees of adaptability. While some models maintain relatively stable performance, others exhibit significant drops when required to decrypt mixed alphanumeric ciphertexts.

For the enhanced prompt template, please refer to \textbf{Figures \ref{tab:enhance1}-\ref{tab:enhance9}}, while more detailed experimental results can be found in \textbf{Table \ref{Enhanced}}.

\begin{figure*}[p]
    \centering
    \begin{exmp}{}{parallel}
        \small
        \setstretch{1.4}  

        \textbf{\#\# Background:}

        \hspace{2em}Decrypt the provided Ciphertext and return the corresponding Plaintext following the given algorithm flow and examples.

        \hspace{2em}Think step by step.

        \hspace{2em}Provide the Plaintext result in the format <result>text</result>, where text is the decrypted text.

        \textbf{\#\# Algorithm Flow:}

        \hspace{2em}Uses the Caesar cipher with a fixed shift of 13 positions. For each letter in the Plaintext, shift it forward by 13 positions in the alphabet to produce the Ciphertext.

        \textbf{\#\# Examples:}

        \hspace{2em}* Example 1:
        
            \hspace{4em}- Ciphertext:
            
                \hspace{6em}Anzr: Rzvyl Wbuafba; Qngr bs Ovegu: Whyl 15, 1990; Cnffcbeg Ahzore: L987654321
                
            \hspace{4em}- Plaintext:
            
                \hspace{6em}Name: Emily Johnson; Date of Birth: July 15, 1990; Passport Number: Y987654321

        \hspace{2em}* Example 2:
        
            \hspace{4em}- Ciphertext:
            
                \hspace{6em}Pnfr Ahzore: 2024-CF-001234; Pnfr Glcr: Gursg/Oernx-Va; Svyvat Qngr: Bpgbore 19, 2024
            
            \hspace{4em}- Plaintext:
            
                \hspace{6em}Case Number: 2024-PS-001234; Case Type: Theft/Break-In; Filing Date: October 19, 2024
                
        \hspace{2em}* Example 3:
        
            \hspace{4em}- Ciphertext:
            
                \hspace{6em}gnk\_vqragvsvre: GKA-2023-NOP456, gnk\_erpbeqf: {lrne: 2023, fgnghf: Cebprffrq, ershaq\_vffhrq: 620.00}
            
            \hspace{4em}- Plaintext:
            
                \hspace{6em}tax\_identifier: TXN-2023-ABC456, tax\_records: {year: 2023, status: Processed, refund\_issued: 620.00}

        \textbf{\#\# Input:}

        \hspace{2em}- Ciphertext:
        
            \hspace{4em}Yrqvn, na Nzrevpna, erfvqrf va Ybf Natryrf.
            
        \hspace{2em}- Plaintext:
        
    \end{exmp}
    \caption{Enhenced Prompt (ROT13 - 3shot - Letter)}
    \label{tab:enhance1}

\end{figure*}

\begin{figure*}[p]
    \centering
    \begin{exmp}{}{parallel}
        \small
        \setstretch{1.4}  

        \textbf{\#\# Background:}

        \hspace{2em}Decrypt the provided Ciphertext and return the corresponding Plaintext following the given algorithm flow and examples.

        \hspace{2em}Think step by step.

        \hspace{2em}Provide the Plaintext result in the format <result>text</result>, where text is the decrypted text.

        \textbf{\#\# Algorithm Flow:}

        \hspace{2em}Uses the Atbash cipher. Each letter in the **Plaintext** is replaced with its reverse counterpart in the alphabet.

        \textbf{\#\# Examples:}

        \hspace{2em}* Example 1:
        
            \hspace{4em}- Ciphertext:
            
                \hspace{6em}Mznv: Vnrob Qlsmhlm; Wzgv lu Yrigs: Qfob 15, 1990; Kzhhklig Mfnyvi: B987654321
                
            \hspace{4em}- Plaintext:
            
                \hspace{6em}Name: Emily Johnson; Date of Birth: July 15, 1990; Passport Number: Y987654321

        \hspace{2em}* Example 2:
        
            \hspace{4em}- Ciphertext:
            
                \hspace{6em}Xzhv Mfnyvi: 2024-KH-001234; Xzhv Gbkv: Gsvug/Yivzp-Rm; Urormt Wzgv: Lxglyvi 19, 2024
            
            \hspace{4em}- Plaintext:
            
                \hspace{6em}Case Number: 2024-PS-001234; Case Type: Theft/Break-In; Filing Date: October 19, 2024
                
        \hspace{2em}* Example 3:
        
            \hspace{4em}- Ciphertext:
            
                \hspace{6em}gzc\_rwvmgrurvi: GCM-2023-ZYX456, gzc\_ivxliwh: {bvzi: 2023, hgzgfh: Kilxvhhvw, ivufmw\_rhhfvw: 620.00}
            
            \hspace{4em}- Plaintext:
            
                \hspace{6em}tax\_identifier: TXN-2023-ABC456, tax\_records: {year: 2023, status: Processed, refund\_issued: 620.00}

        \textbf{\#\# Input:}

        \hspace{2em}- Ciphertext:
        
            \hspace{4em}Ovwrz, zm Znvirxzm, ivhrwvh rm Olh Zmtvovh.
            
        \hspace{2em}- Plaintext:
        
    \end{exmp}
    \caption{Enhenced Prompt (Atbash - 3shot - Letter)}
    \label{tab:enhance2}

\end{figure*}

\begin{figure*}[p]
    \centering
    \begin{exmp}{}{parallel}
        \small
        \setstretch{1.4}  

        \textbf{\#\# Background:}

        \hspace{2em}Decrypt the provided Ciphertext and return the corresponding Plaintext following the given algorithm flow and examples.

        \hspace{2em}Think step by step.

        \hspace{2em}Provide the Plaintext result in the format <result>text</result>, where text is the decrypted text.

        \textbf{\#\# Algorithm Flow:}

        \hspace{2em}Uses the Polybius cipher. Each letter in the **Plaintext** is mapped to a pair of coordinates in the Polybius square, forming the **Ciphertext**.

        \textbf{\#\# Examples:}

        \hspace{2em}* Example 1:
        
            \hspace{4em}- Ciphertext:
            
                \hspace{6em}32 11 31 15 :   15 31 23 26 51   24 33 22 32 41 33 32 ;   
                14 11 42 15   33 16   12 23 36 42 22 :   24 43 26 51   1 5 ,   1 9 9 0 ;   
                34 11 41 41 34 33 36 42   32 43 31 12 15 36 :   51 9 8 7 6 5 4 3 2 1
            
            \hspace{4em}- Plaintext:
            
                \hspace{6em}Name: Emily Johnson; Date of Birth: July 15, 1990; Passport Number: Y987654321

        \hspace{2em}* Example 2:
        
            \hspace{4em}- Ciphertext:
            
                \hspace{6em}13 11 41 15   32 43 31 12 15 36 :   2 0 2 4 - 34 41 - 0 0 1 2 3 4 ;   
                13 11 41 15   42 51 34 15 :   42 22 15 16 42 / 12 36 15 11 25 - 23 32 ;   
                16 23 26 23 32 21   14 11 42 15 :   33 13 42 33 12 15 36   1 9 ,   2 0 2 4
            
            \hspace{4em}- Plaintext:
            
                \hspace{6em}Case Number: 2024-PS-001234; Case Type: Theft/Break-In; Filing Date: October 19, 2024
                
        \hspace{2em}* Example 3:
        
            \hspace{4em}- Ciphertext:
            
                \hspace{6em}42 11 46 $\_$ 23 14 15 32 42 23 16 23 15 36 :   
                42 46 32 $\_$ 2 0 2 3 - 11 12 13 4 5 6 ,   
                42 11 46 $\_$ 36 15 13 33 36 14 41 :   
                { 51 15 11 36 :   2 0 2 3 ,   41 42 11 42 43 41 :   34 36 33 13 15 41 41 15 14 ,   
                36 15 16 43 32 14 $\_$ 23 41 41 43 15 14 :   6 2 0 . 0 0 }
            
            \hspace{4em}- Plaintext:
            
                \hspace{6em}tax\_identifier: TXN-2023-ABC456, tax\_records: {year: 2023, status: Processed, refund\_issued: 620.00}

        \textbf{\#\# Input:}

        \hspace{2em}- Ciphertext:
        
            \hspace{4em}26 15 14 23 11 ,   11 32   11 31 15 36 23 13 11 32 ,   
            36 15 41 23 14 15 41   23 32   26 33 41   11 32 21 15 26 15 41 .
            
        \hspace{2em}- Plaintext:
        
    \end{exmp}
    \caption{Enhanced Prompt (Polybius - 3shot - Letter)}
    \label{tab:enhance3}

\end{figure*}

\begin{figure*}[p]
    \centering
    \begin{exmp}{}{parallel}
        \small
        \setstretch{1.4}  

        \textbf{\#\# Background:}

        \hspace{2em}Decrypt the provided Ciphertext and return the corresponding Plaintext following the given algorithm flow and examples.

        \hspace{2em}Think step by step.

        \hspace{2em}Provide the Plaintext result in the format <result>text</result>, where text is the decrypted text.

        \textbf{\#\# Algorithm Flow:}

        \hspace{2em}Uses the Vigenère cipher. Each letter in the **Plaintext** is shifted by the corresponding letter in the **Key** to produce the **Ciphertext**.

        \textbf{\#\# Examples:}

        \hspace{2em}* Example 1:
        
            \hspace{4em}- Ciphertext:
            
                \hspace{6em}Ncxe: Eotla Jqsnuzn; Dcee zf Miteh: Jwwy 15, 1990; Pcdsrzrv Nwxbgc: J987654321
                
            \hspace{4em}- Plaintext:
            
                \hspace{6em}Name: Emily Johnson; Date of Birth: July 15, 1990; Passport Number: Y987654321

        \hspace{2em}* Example 2:
        
            \hspace{4em}- Ciphertext:
            
                \hspace{6em}Ccde Yuomet: 2024-PU-001234; Naup Vjpg: Vsehe/Dcecv-Ky; Qintni Dcee: Oeeodpr 19, 2024
            
            \hspace{4em}- Plaintext:
            
                \hspace{6em}Case Number: 2024-PS-001234; Case Type: Theft/Break-In; Filing Date: October 19, 2024
                
        \hspace{2em}* Example 3:
        
            \hspace{4em}- Ciphertext:
            
                \hspace{6em}tci\_koepeihtet: VIN-2023-CMC456, tci\_tpcqcdu: {jecc: 2023, dtceuu: Rcoepsupd, rgqupo\_kdswpd: 620.00}
            
            \hspace{4em}- Plaintext:
            
                \hspace{6em}tax\_identifier: TXN-2023-ABC456, tax\_records: {year: 2023, status: Processed, refund\_issued: 620.00}

        \textbf{\#\# Input:}

        \hspace{2em}- Ciphertext:
        
            \hspace{4em}Lgoic, cy Cxettccy, ceutdgd ky Nzs Lniplgd.
            
        \hspace{2em}- Plaintext:
        
    \end{exmp}
    \caption{Enhanced Prompt (Vigenère - 3shot - Letter)}
    \label{tab:enhance4}

\end{figure*}

\begin{figure*}[p]
    \centering
    \begin{exmp}{}{parallel}
        \small
        \setstretch{1.4}  

        \textbf{\#\# Background:}

        \hspace{2em}Decrypt the provided Ciphertext and return the corresponding Plaintext following the given algorithm flow and examples.

        \hspace{2em}Think step by step.

        \hspace{2em}Provide the Plaintext result in the format <result>text</result>, where text is the decrypted text.

        \textbf{\#\# Algorithm Flow:}

        \hspace{2em}Reverses the **Plaintext** to create the **Ciphertext**.

        \textbf{\#\# Examples:}

        \hspace{2em}* Example 1:
        
            \hspace{4em}- Ciphertext:
            
                \hspace{6em}123456789Y :rebmuN tropssaP ;0991 ,51 yluJ :htriB fo etaD ;nosnhoJ ylimE :emaN
                
            \hspace{4em}- Plaintext:
            
                \hspace{6em}Name: Emily Johnson; Date of Birth: July 15, 1990; Passport Number: Y987654321

        \hspace{2em}* Example 2:
        
            \hspace{4em}- Ciphertext:
            
                \hspace{6em}4202 ,91 rebotcO :etaD gniliF ;nI-kaerB/tfehT :epyT esaC ;432100-SP-4202 :rebmuN esaC
            
            \hspace{4em}- Plaintext:
            
                \hspace{6em}Case Number: 2024-PS-001234; Case Type: Theft/Break-In; Filing Date: October 19, 2024
                
        \hspace{2em}* Example 3:
        
            \hspace{4em}- Ciphertext:
            
                \hspace{6em}\}00.026 :deussi\_dnufer ,dessecorP :sutats ,3202 :raey\{ :sdrocer\_xat ,654CBA-3202-NXT :reifitnedi\_xat
            
            \hspace{4em}- Plaintext:
            
                \hspace{6em}tax\_identifier: TXN-2023-ABC456, tax\_records: {year: 2023, status: Processed, refund\_issued: 620.00}

        \textbf{\#\# Input:}

        \hspace{2em}- Ciphertext:
        
            \hspace{4em}.selegnA soL ni sediser ,naciremA na ,aideL
            
        \hspace{2em}- Plaintext:
        
    \end{exmp}
    \caption{Enhenced Prompt (Reverse - 3shot - Letter)}
    \label{tab:enhance5}

\end{figure*}

\begin{figure*}[p]
    \centering
    \begin{exmp}{}{parallel}
        \small
        \setstretch{1.4}  

        \textbf{\#\# Background:}

        \hspace{2em}Decrypt the provided Ciphertext and return the corresponding Plaintext following the given algorithm flow and examples.

        \hspace{2em}Think step by step.

        \hspace{2em}Provide the Plaintext result in the format <result>text</result>, where text is the decrypted text.

        \textbf{\#\# Algorithm Flow:}

        \hspace{2em}For each pair of letters in the **Plaintext**, their positions are swapped to produce the **Ciphertext**. If the number of letters is odd, the last letter remains in its original position.

        \textbf{\#\# Examples:}

        \hspace{2em}* Example 1:
        
            \hspace{4em}- Ciphertext:
            
                \hspace{6em}aNem :mEli yoJnhos;nD ta efoB riht :uJyl1 ,51 99;0P sapsro tuNbmre :9Y78563412
                
            \hspace{4em}- Plaintext:
            
                \hspace{6em}Name: Emily Johnson; Date of Birth: July 15, 1990; Passport Number: Y987654321

        \hspace{2em}* Example 2:
        
            \hspace{4em}- Ciphertext:
            
                \hspace{6em}aCesN mueb:r2 20-4SP0-1032;4C sa eyTep :hTfe/trBae-knI ;iFilgnD ta:eO tcbore1 ,92 204
            
            \hspace{4em}- Plaintext:
            
                \hspace{6em}Case Number: 2024-PS-001234; Case Type: Theft/Break-In; Filing Date: October 19, 2024
                
        \hspace{2em}* Example 3:
        
            \hspace{4em}- Ciphertext:
            
                \hspace{6em}at\_xdineitifre :XT-N0232A-CB54,6t xar\_cerosd :y{ae:r2 20,3s atut:sP orecssde ,erufdni\_sseu:d6 020.}0
            
            \hspace{4em}- Plaintext:
            
                \hspace{6em}tax\_identifier: TXN-2023-ABC456, tax\_records: {year: 2023, status: Processed, refund\_issued: 620.00}

        \textbf{\#\# Input:}

        \hspace{2em}- Ciphertext:
        
            \hspace{4em}eLid,aa  nmArecina ,erised sniL soA gnlese.
            
        \hspace{2em}- Plaintext:
        
    \end{exmp}
    \caption{Enhenced Prompt (SwapPairs - 3shot - Letter)}
    \label{tab:enhance6}

\end{figure*}

\begin{figure*}[p]
    \centering
    \begin{exmp}{}{parallel}
        \small
        \setstretch{1.4}  

        \textbf{\#\# Background:}

        \hspace{2em}Decrypt the provided Ciphertext and return the corresponding Plaintext following the given algorithm flow and examples.

        \hspace{2em}Think step by step.

        \hspace{2em}Provide the Plaintext result in the format <result>text</result>, where text is the decrypted text.

        \textbf{\#\# Algorithm Flow:}

        \hspace{2em}This encryption method converts each letter of the **Plaintext** into two letters in the **Ciphertext**, such that the average of their ASCII values equals the ASCII value of the original letter.

        \textbf{\#\# Examples:}

        \hspace{2em}* Example 1:
        
            \hspace{4em}- Ciphertext:
            
                \hspace{6em}MOaalndf: DFlnhjkmxz IKnpgimortnpmo; CEaasudf npeg AChjqssugi: IKtvkmxz 15, 1990; OQaartrtoqnpqssu MOtvlnacdfqs: XZ987654321
                
            \hspace{4em}- Plaintext:
            
                \hspace{6em}Name: Emily Johnson; Date of Birth: July 15, 1990; Passport Number: Y987654321

        \hspace{2em}* Example 2:
        
            \hspace{4em}- Ciphertext:
            
                \hspace{6em}BDaartdf MOtvlnacdfqs: 2024-OQRT-001234; BDaartdf SUxzoqdf: SUgidfegsu/ACqsdfaajl-HJmo; EGhjkmhjmofh CEaasudf: NPbdsunpacdfqs 19, 2024
            
            \hspace{4em}- Plaintext:
            
                \hspace{6em}Case Number: 2024-PS-001234; Case Type: Theft/Break-In; Filing Date: October 19, 2024
                
        \hspace{2em}* Example 3:
        
            \hspace{4em}- Ciphertext:
            
                \hspace{6em}suaawy\_hjcedfmosuhjeghjdfqs: SUWYMO-2023-AAACBD456, suaawy\_qsdfbdnpqscert: {xzdfaaqs: 2023, rtsuaasutvrt: OQqsnpbddfrtrtdfce, qsdfegtvmoce\_hjrtrttvdfce: 620.00}
            
            \hspace{4em}- Plaintext:
            
                \hspace{6em}tax\_identifier: TXN-2023-ABC456, tax\_records: {year: 2023, status: Processed, refund\_issued: 620.00}

        \textbf{\#\# Input:}

        \hspace{2em}- Ciphertext:
        
            \hspace{4em}KMdfcehjaa, aamo AAlndfqshjbdaamo, qsdfrthjcedfrt hjmo KMnprt AAmofhdfkmdfrt.
            
        \hspace{2em}- Plaintext:
        
    \end{exmp}
    \caption{Enhenced Prompt (DualAvgCode - 3shot - Letter)}
    \label{tab:enhance7}

\end{figure*}

\begin{figure*}[p]
    \centering
    \begin{exmp}{}{parallel}
        \small
        \setstretch{1.4}  

        \textbf{\#\# Background:}

        \hspace{2em}Decrypt the provided Ciphertext and return the corresponding Plaintext following the given algorithm flow and examples.

        \hspace{2em}Think step by step.

        \hspace{2em}Provide the Plaintext result in the format <result>text</result>, where text is the decrypted text.

        \textbf{\#\# Algorithm Flow:}

        \hspace{2em}For each letter in the **Plaintext**:  
        \hspace{3em}- If the ASCII value is even, add 1 to it to get the corresponding character in the **Ciphertext**.  
        \hspace{3em}- If the ASCII value is odd, subtract 1 to get the new character in the **Ciphertext**.

        \textbf{\#\# Examples:}

        \hspace{2em}* Example 1:
        
            \hspace{4em}- Ciphertext:
            
                \hspace{6em}Ozld: Dlhmx Kniorno; Ezud ng Chsui: Ktmx 15, 1990; Qzrrqnsu Otlcds: X987654321
                
            \hspace{4em}- Plaintext:
            
                \hspace{6em}Name: Emily Johnson; Date of Birth: July 15, 1990; Passport Number: Y987654321

        \hspace{2em}* Example 2:
        
            \hspace{4em}- Ciphertext:
            
                \hspace{6em}Bzrd Otlcds: 2024-QR-001234; Bzrd Uxqd: Uidgu/Csdzj-Ho; Ghmhof Ezud: Nbuncds 19, 2024
            
            \hspace{4em}- Plaintext:
            
                \hspace{6em}Case Number: 2024-PS-001234; Case Type: Theft/Break-In; Filing Date: October 19, 2024
                
        \hspace{2em}* Example 3:
        
            \hspace{4em}- Ciphertext:
            
                \hspace{6em}uzy\_hedouhghds: UYO-2023-ZCB456, uzy\_sdbnser: {xdzs: 2023, ruzutr: Qsnbdrrde, sdgtoe\_hrrtde: 620.00}
            
            \hspace{4em}- Plaintext:
            
                \hspace{6em}tax\_identifier: TXN-2023-ABC456, tax\_records: {year: 2023, status: Processed, refund\_issued: 620.00}

        \textbf{\#\# Input:}

        \hspace{2em}- Ciphertext:
        
            \hspace{4em}Mdehz, zo Zldshbzo, sdrhedr ho Mnr Zofdmdr.
            
        \hspace{2em}- Plaintext:
        
    \end{exmp}
    \caption{Enhenced Prompt (ParityShift - 3shot - Letter)}
    \label{tab:enhance8}

\end{figure*}

\begin{figure*}[p]
    \centering
    \begin{exmp}{}{parallel}
        \small
        \setstretch{1.4}  

        \textbf{\#\# Background:}

        \hspace{2em}Decrypt the provided Ciphertext and return the corresponding Plaintext following the given algorithm flow and examples.

        \hspace{2em}Think step by step.

        \hspace{2em}Provide the Plaintext result in the format <result>text</result>, where text is the decrypted text.

        \textbf{\#\# Algorithm Flow:}

        \hspace{2em}The algorithm splits the **Plaintext** into words based on spaces. Each word is then individually encrypted using the Caesar cipher, resulting in the **Ciphertext**.

        \textbf{\#\# Examples:}

        \hspace{2em}* Example 1:
        
            \hspace{4em}- Ciphertext:
            
                \hspace{6em}e:Nam lyEmi nson;Joh eDat fo th:Bir yJul 15, 0;199 sportPas ber:Num 7654321Y98
                
            \hspace{4em}- Plaintext:
            
                \hspace{6em}Name: Emily Johnson; Date of Birth: July 15, 1990; Passport Number: Y987654321

        \hspace{2em}* Example 2:
        
            \hspace{4em}- Ciphertext:
            
                \hspace{6em}eCas ber:Num 4-PS-001234;202 eCas e:Typ ft/Break-In;The ingFil e:Dat oberOct 19, 4202
            
            \hspace{4em}- Plaintext:
            
                \hspace{6em}Case Number: 2024-PS-001234; Case Type: Theft/Break-In; Filing Date: October 19, 2024
                
        \hspace{2em}* Example 3:
        
            \hspace{4em}- Ciphertext:
            
                \hspace{6em}\_identifier:tax -2023-ABC456,TXN \_records:tax ar:{ye 3,202 tus:sta cessed,Pro und\_issued:ref .00}620
            
            \hspace{4em}- Plaintext:
            
                \hspace{6em}tax\_identifier: TXN-2023-ABC456, tax\_records: {year: 2023, status: Processed, refund\_issued: 620.00}

        \textbf{\#\# Input:}

        \hspace{2em}- Ciphertext:
        
            \hspace{4em}ia,Led na rican,Ame idesres ni Los eles.Ang
            
        \hspace{2em}- Plaintext:
        
    \end{exmp}
    \caption{Enhenced Prompt (WordShift - 3shot - Letter)}
    \label{tab:enhance9}

\end{figure*}

\begin{table*}[h!]
    \centering
    \caption{Results on CipherBank(Enhanced Prompt)}
    \label{Enhanced}
    \resizebox{\textwidth}{!}
    {
    \begin{tabular}{lcccccccccccc}
        \toprule
        \multirow{2}{*}{\textbf{Model}} & \multicolumn{4}{c}{\textbf{Substitution Ciphers}} & \multicolumn{2}{c}{\textbf{Transposition Ciphers}}  & \multicolumn{3}{c}{\textbf{Custom Ciphers}}  & \multicolumn{1}{c}{\textbf{Cipher Score}}  \\
        \cmidrule(lr){2-5} \cmidrule(lr){6-7} \cmidrule(lr){8-10} \cmidrule(lr){11-11} 
        &Rot13 & Atbash & Polybius & Vigenère & Reverse & SwapPairs  & DualAvgCode & ParityShift & WordShift & Accuracy$_{\text{avg}}$\\
        \midrule
        \rowcolor[HTML]{f8e2d1}
        \multicolumn{11}{c}{\textit{Open-source Chat Models}} \\ \midrule
        Mixtral-8x22B-v0.1 &0.76& 0 &0 & 0 & 0.38 &0 & 2.67 & 0.38 & 0.38 & 0.51 \\
        Qwen2.5-72B-Instruct &12.60&9.16 & 0 &0 & 0 & 0 & 2.29 & 0.38 & 1.53 & 2.88\\  
        Llama-3.1-70B-Instruct &2.67& 1.15 & 0 & 0 & 1.53 & 0.38 &1.15 & 0 & 0& 0.76 \\ 
        Llama-3.3-70B-Instruct &4.58&1.53 & 0 & 0.38 &1.15 & 0 & 1.15 & 0 & 0 & 0.98\\ 
        DeepSeek-V3 &41.60 &27.86& 0.38 & 0.38&\underline{65.95} &5.34 &12.66 &0.76 & 5.17 &17.79\\  \midrule
        \rowcolor[HTML]{fdf3d0}
        \multicolumn{11}{c}{\textit{Closed-source Models}} \\ \midrule
        GPT-4o-mini-2024-07-18 & 21.76 & 19.08 & 0 & 0.38 & 4.39 & 0 & 0 & 0 & 0 & 5.07\\
        GPT-4o-2024-08-06 & 45.42 & 24.05 &0 &\underline{0.76}&51.53&8.40&1.91 &1.15&10.31 & 15.95\\
        GPT-4o-2024-11-20  & 45.42 & 41.98 & 0 & 0 &  53.63 & 8.02 & 3.82 &1.15& 9.54 & 18.17\\
        gemini-1.5-pro &63.69 & 5.73 & 0.76 &  0.38 & 14.12 & 2.67 & 0.38 & 1.91 & \underline{10.69} & 11.15\\
        gemini-2.0-flash-exp & 45.04 & 22.90 & 2.29 & 0.38 & 46.56 & 4.58 & 3.82 & 0 &1.15 & 14.08 \\
        Claude-Sonnet-3.5-1022 & \textbf{92.75}  &\underline{82.06} &\textbf{78.24}&\textbf{2.48 }&\textbf{79.39} & 9.73 &2.48 & \underline{62.02} & \textbf{44.85 }& \underline{50.44}\\
        \midrule
        \rowcolor[HTML]{d6ecda}
        \multicolumn{11}{c}{\textit{Reasoning Models}} \\ \midrule
        QwQ-32B-Preview & 1.91 & 3.05 & 2.67 & 0 & 0 & 0  & 2.67 & 0.38 & 0.38 & 1.23\\
        DeepSeek-R1 & \underline{88.37} &\textbf{ 86.54} &\underline{72.73} & \underline{0.76} & 46.96 & \textbf{75.01 }& \textbf{73.17} &\textbf{ 74.42} & 1.51 &\textbf{57.72}\\ 
        gemini-2.0-flash-thinking & 37.98 & 19.09 & 10.50 & 0 & 55.34 & 4.96 & 4.77 & 0.38 & 6.11&  15.46   \\
        o1-mini-2024-09-12 & 54.20 & 72.14 & 50.0 & \underline{0.76} & 11.07 & \underline{18.70} &  \underline{47.33} & 49.62 & 7.25 & 34.56\\
        \midrule
        \bottomrule
    \end{tabular}%
    }
\end{table*}

\subsection{Impact of Plaintext Source on Decryption Performance}

To assess how plaintext characteristics influence decryption performance, we compare results on synthetically generated privacy-sensitive data versus externally sourced structured text (e.g., quotes from Shakespeare’s works). The structured text exhibits greater linguistic familiarity, while the privacy-sensitive data represents real-world encryption needs, lacking inherent semantic patterns.  

As shown in\textbf{ Table~\ref{table_common_synthetic}} and \textbf{Table~\ref{table_common_structured},} models generally perform better on structured text, suggesting that they leverage linguistic priors rather than strictly following decryption rules. When encountering encrypted text with recognizable patterns, models tend to shortcut reasoning, aligning decoded fragments with plausible linguistic structures instead of strictly adhering to learned transformation rules. Conversely, for less structured, domain-specific text, models struggle to infer decryption patterns, reinforcing the advantage of CipherBank’s privacy-sensitive dataset, which forces models to engage in independent reasoning rather than rely on pretraining biases.

\begin{table*}[h!]
    \centering
    \caption{Decryption Performance on Privacy-Sensitive Data}
    \label{table_common_synthetic}
    \resizebox{\textwidth}{!}{%
    \begin{tabular}{lccccccccccc}
        \toprule
        \textbf{Model}  & Rot13 & Atbash & Polybius & Vigenère & Reverse & Swap & DualAvgCode  & ParityShift & WordShift & Accuracy$_{\text{avg}}$\\
        \midrule
        \multicolumn{11}{c}{\textit{Open-source Models}} \\ \midrule
        DeepSeek-V3 &  24.34 & 15.64 & 15.70 & 0 & 33.72 & 3.51 & 0 & 4.35 & 15.64  & 12.54 \\ 
        DeepSeek-R1 & 57.88 & 71.02 & 71.55  &4.35 &33.57 & 4.35 & 0 & 12.71 & 8.70  & 29.35 \\  
        \midrule
        \multicolumn{11}{c}{\textit{Closed-source Models}} \\ \midrule
        GPT-4o-2024-11-20 & 21.74 & 21.74 & 0 & 0 & 30.43 & 8.70 & 0 & 0 & 13.04 & 10.63 \\
        Gemini-2.0-Flash-Exp & 47.83 & 4.35 & 4.35 & 0 & 52.17 & 0 & 4.35 & 4.35 & 13.04 & 14.49 \\
        Claude-Sonnet-3.5-1022 & 86.96 & 78.26 & 65.22 & 4.35  & 91.30 & 13.04 & 4.35 & 52.17 & 47.83 & 49.28 \\
        Gemini-2.0-Flash-Thinking & 39.13 & 4.35 & 0 & 0 & 60.87 & 0 & 0 & 4.35 & 30.43 & 15.46 \\
        o1-Mini-2024-09-12 & 60.87 & 86.96 & 69.57 & 0 & 8.70 & 0 & 13.04 & 17.39 & 4.35 & 28.99 \\
        \midrule
        \bottomrule
    \end{tabular}%
    }
\end{table*}

\begin{table*}[h!]
    \centering
    \caption{Decryption Performance on Structured Text}
    \label{table_common_structured}
    \resizebox{\textwidth}{!}{%
    \begin{tabular}{lccccccccccc}
        \toprule
        \textbf{Model}  & Rot13 & Atbash & Polybius & Vigenère & Reverse & SwapPair & DualAvgCode  & ParityShift & WordShift & Accuracy$_{\text{avg}}$\\
        \midrule
        \multicolumn{11}{c}{\textit{Open-source Models}} \\ \midrule
        DeepSeek-V3 &  76.12 & 24.03 & 15.70 & 0 & 52.17 & 29.40 & 0 & 12.71 & 55.13  & 29.47 \\ 
        DeepSeek-R1 & 84.51 & 85.04 & 100  &7.59 &79.10 & 8.70 & 8.70 & 15.64 & 30.43  & 46.63 \\  
        \midrule
        \multicolumn{11}{c}{\textit{Closed-source Models}} \\ \midrule
        GPT-4o-2024-11-20 & 78.26 & 39.13 & 4.35 & 0 & 86.96 & 21.74 & 0 & 4.35 & 43.48 & 30.92 \\
        Gemini-2.0-Flash-Exp & 86.96 & 13.04 & 4.35 & 0 & 86.96 & 8.70 & 0 & 17.39 & 43.48 & 28.99 \\
        Claude-Sonnet-3.5-1022 & 91.30 & 95.65 & 95.65 & 4.35  & 100 & 52.17 & 8.70 & 78.26 & 95.65 & 69.08 \\
        Gemini-2.0-Flash-Thinking & 86.96 & 13.04 & 8.70 & 0 & 69.57 & 17.39 & 0 & 0 & 52.17 & 27.54 \\
        o1-Mini-2024-09-12 & 82.61 & 95.65 & 78.26 & 0 & 60.87 & 4.35 & 13.04 & 17.39 & 43.48 & 43.96 \\
        \midrule
        \bottomrule
    \end{tabular}%
    }
\end{table*}

\section{Error Analysis}
\subsection{Error Classification}
\label{error1}

This section defines the error categories observed in model decryption outputs. These classifications help identify systematic failure patterns and provide insights into how models approach cryptographic reasoning.  

\begin{itemize}
    \item (A) Omission/Insertion: The model output contains missing or extra characters, words, or punctuation compared to the reference plaintext. These errors indicate incomplete decryption or unintended modifications, leading to partial but inaccurate results.  

    \item (B) Name Decryption Error: The decryption result is correct except for the name part, which remains incorrect or partially distorted. This suggests challenges in handling named entities, possibly due to memorization effects or entity-based biases.  

    \item (C) Semantic Inference: The model makes errors based on semantic reasoning rather than strictly following decryption rules. Instead of decoding symbols precisely, the model hallucinates plausible but incorrect outputs that fit the general meaning of the sentence. This indicates a tendency to prioritize linguistic coherence over strict decryption fidelity.  

    \item (D) Reorganization: The output preserves the exact meaning of the reference plaintext but rearranges the sentence structure. This suggests that the model prioritizes fluency over strict character-level fidelity, leading to errors in cryptographic tasks where precision is essential.  

    \item (E) Reasoning Failure: The model output is significantly different from the reference, and decryption is essentially unsuccessful. This suggests a fundamental failure in identifying encryption patterns, leading to outputs that bear little resemblance to the expected plaintext. This category includes cases where the model fails to infer transformation rules or apply correct decryption strategies.  

    \item (F) Other: Miscellaneous errors that do not fit into the defined categories.
\end{itemize}  

This classification framework provides a structured approach to analyzing decryption errors, helping to pinpoint systematic weaknesses and guide future improvements in cryptographic reasoning models.

\subsection{Examples of Different Error Types}
\label{error2}

\begin{figure*}[p]
    \centering
    \begin{exmp}{Error Type: Omission/Insertion}{parallel}
        \small
        \setstretch{1.4}  

        \textbf{Example 1:}

        \textbf{Predictions:} Card Number: ID 1245-6789-0123  

        \textbf{References:} Clark holds the ID Card Number 1245-6789-0123.  

        \vspace{2mm} 

        \textbf{Example 2:}

        \textbf{Predictions:} Card Number: ID 1245-6789-0123  

        \textbf{References:} Clark holds the ID Card Number 1245-6789-0123.  

        \vspace{2mm} 

        \textbf{Example 3:}

        \textbf{Predictions:} Salary Amount: \$67,000; Pay Date: 2023-10-25  

        \textbf{References:} Salary Amount: \$67,000, Pay Date: 2023-10-25.  

    \end{exmp}
\end{figure*}

\begin{figure*}[p]
    \centering
    \begin{exmp}{Error Type: Name Decryption Error}{parallel}
        \small
        \setstretch{1.4}  

        \textbf{Example 1:}  

        \textbf{Predictions:} Learn, an American, inside on Los Angeles.  

        \textbf{References:} Ledia, an American, resides in Los Angeles.  

        \vspace{2mm} 

        \textbf{Example 2:}  

        \textbf{Predictions:}  
        Individual ID: A1234567; Name: John Doe; Age: 34;  
        Gender Identity: Cisgender 16k11.2 Location, Country State Citizenship.  

        \textbf{References:}  
        Individual ID: A1234567; Name: Jane Doe; Age: 34;  
        Genetic Testing: Chromosome 16p11.2 Deletion, Celiac Disease Predisposition.  

        \vspace{2mm} 

        \textbf{Example 3:}  

        \textbf{Predictions:} Handed lost the passport Number A12345678.  

        \textbf{References:} Dean holds the passport number A12345678.  

    \end{exmp}
\end{figure*}

\begin{figure*}[p]
    \centering
    \begin{exmp}{Error Type: Semantic Inference}{parallel}
        \small
        \setstretch{1.4}  

        \textbf{Example 1:}

        \textbf{Predictions:} Jessica Brown, Bachelor of Biology, GPA: 3.9, Graduated 2023, Skills: Genetics, Microbiology, Ecology, Bioinformatics.  

        \textbf{References:} Jessica Brown, Bachelors in Biology, GPA: 3.9, Graduated 2023, Courses: Genetics, Microbiology, Ecology, Biochemistry.  

        \vspace{2mm} 

        \textbf{Example 2:}

        \textbf{Predictions:} Patent-pending design specification PR2023\_KURITY, Company Z, including batch production requirements.  

        \textbf{References:} Patent-pending design specification PR2023\_KURITY, Company Z, including batch performance analysis.  

        \vspace{2mm} 

        \textbf{Example 3:}

        \textbf{Predictions:} Office ID: P12345, Name: John, Position: Manager, Department: Finance.  

        \textbf{References:} Officer ID: P12345, Name: John, Position: Sergeant, Department: Homicide.  

    \end{exmp}
\end{figure*}

\begin{figure*}[p]
    \centering
    \begin{exmp}{Error Type: Reorganization}{parallel}
        \small
        \setstretch{1.4}  
        
        \textbf{Example 1:}
        
        \textbf{Predictions:} Name: Ledia; Nationality: American; Address: Los Angeles  
        
        \textbf{References:} Ledia, an American, resides in Los Angeles.  
        
        \vspace{2mm} 

        \textbf{Example 2:}
        
        \textbf{Predictions:} Passport Number: A12345678; Holder Name: Dean  
        
        \textbf{References:} Dean holds the passport number A12345678.
        
        \vspace{2mm} 

        \textbf{Example 3:}
        
        \textbf{Predictions:} Name: Peter; Date of Birth: April 23, 1985; Passport Number: X123456789  
        
        \textbf{References:} Peter was born on April 23, 1985, and carries a passport with the number X123456789.

    \end{exmp}
\end{figure*}

\begin{figure*}[p]
    \centering
    \begin{exmp}{Error Type: Reasoning Failure}{parallel}
        \small
        \setstretch{1.4}  

        \textbf{Example 1:}

        \textbf{Predictions:} Address: 123 Main St, Apt 4B, New York, NY, Zip Code: 10001, Phone: 555-1234, Unit: 101, Floor: 2.  

        \textbf{References:} Officer: Lisa Grant, Sergeant, Downtown Precinct, Patrol Car: VG2301, Equipment: Radar Gun, Model: RGX501, Weapon: Taser X2.  

        \vspace{2mm} 

        \textbf{Example 2:}

        \textbf{Predictions:} Welcome, once more, securely within.  

        \textbf{References:} Ledia, an American, resides in Los Angeles.  

        \vspace{2mm} 

        \textbf{Example 3:}

        \textbf{Predictions:} Passport ID: 123456789; Expiry Date: Emily, 38, Issued By Authority, Renewal By Agency. Valid Until 2025, Expiry Passport.  

        \textbf{References:} Patient ID: MRN456789; Medical History: Emily, 38, diagnosed with asthma, treated with inhalers. Allergic to dust mites, pollen.  

    \end{exmp}
\end{figure*}

\begin{figure*}[p]
    \centering
    \begin{exmp}{Error Type: Other}{parallel}
        \small
        \setstretch{1.4}  

        \textbf{Example 1:}

        \textbf{Predictions:} CookieID12345 maintain login status for UserID98765 on www.example.com, facilitating seamless access. Analyzing records UserID98765's engagement, deducting 500 page views and a click-through rate of 4.5\% across the session.  

        \textbf{References:} CookieID12345 maintains login status for UserID98765 on www.example.com, facilitating seamless access. Analytics tracks UserID98765's engagement, documenting 500 page views and a click-through rate of 4.5\% across the session.  

        \vspace{2mm} 

        \textbf{Example 2:}

        \textbf{Predictions:} Code: Coordinates: Latitude Longitude: 38.251° N, -85.754° W, Latitude Longitude: 34.091° N, -118.493° W.  

        \textbf{References:} Base Distribution: North Plains Base: 38.251° N, -85.754° W, East Valley Site: 34.091° N, -118.493° W.  

        \vspace{2mm} 

        \textbf{Example 3:}

        \textbf{Predictions:} Name: Alex Smith; Salary: \$87,500; Pay Frequency: Biweekly; Position: Software Developer; Employee ID: EID-257846; Department: IT.  

        \textbf{References:} Name: Alex Smith, Salary: \$87,500, Pay Frequency: Biweekly, Position: Software Developer, Employee ID: EID-257846, Department: IT.  

    \end{exmp}
\end{figure*}

To further illustrate the types of decryption errors encountered in our evaluation, we provide concrete examples corresponding to each error category. These cases demonstrate how models fail in various aspects of decryption, including omission/insertion, name decryption errors, semantic inference, reorganization, reasoning failures, and other anomalies. \textbf{Example D.1 - D6} showcase representative examples of each error type.

\subsection{Detailed Error Distribution Tables}
\label{error3} 

\begin{table*}[h!]
\centering
\caption{Error Type Percentages for Different Algorithms in Claude-Sonnet-3.5-1022 Model}
\label{tab:error_1}
\resizebox{\textwidth}{!}{
\begin{tabular}{c|cccccc}
\toprule
\multirow{2}{*}{Algorithm} & \multicolumn{6}{c}{Error Types} \\
& Omission/Insertion & Name Decryption Error & Semantic Inference & Reorganization & Reasoning Failure & Other \\
\midrule
Rot13 & 33.33 & 51.85 & 0.00 & 11.11 & 3.70 & 0.00 \\
Atbash & 15.79 & 78.95 & 0.00 & 3.51 & 0.00 & 1.75 \\
Polybius & 42.62 & 45.90 & 0.00 & 11.48 & 0.00 & 0.00 \\
Vigenère & 2.73 & 32.42 & 5.08 & 3.52 & 56.25 & 0.00 \\
Reverse & 39.24 & 48.10 & 0.00 & 5.06 & 6.33 & 1.27 \\
SwapPairs & 15.98 & 38.52 & 2.05 & 2.87 & 38.11 & 2.46 \\
DualAvgCode & 6.88 & 39.68 & 8.50 & 2.43 & 41.30 & 1.21 \\
ParityShift & 19.79 & 70.83 & 4.17 & 3.12 & 2.08 & 0.00 \\
WordShift & 51.95 & 22.08 & 2.60 & 8.44 & 12.34 & 2.60 \\
\bottomrule
\end{tabular}
}
\end{table*}

\begin{table*}[h!]
\centering
\caption{Error Type Percentages for Different Algorithms in DeepSeek-R1 Model}
\label{tab:error_2}
\resizebox{\textwidth}{!}{
\begin{tabular}{c|cccccc}
\toprule
\multirow{2}{*}{Algorithm} & \multicolumn{6}{c}{Error Types} \\
& Omission/Insertion & Name Decryption Error & Semantic Inference & Reorganization & Reasoning Failure & Other \\
\midrule
Rot13 & 40.00 & 30.00 & 4.29 & 21.43 & 1.43 & 2.86 \\
Atbash & 42.59 & 24.07 & 0.93 & 29.63 & 0.00 & 2.78 \\
Polybius & 48.63 & 17.12 & 0.68 & 21.92 & 8.90 & 2.74 \\
Vigenère & 4.60 & 18.01 & 2.68 & 2.30 & 71.65 & 0.77 \\
Reverse & 25.64 & 19.66 & 1.71 & 45.30 & 6.41 & 1.28 \\
SwapPairs & 9.20 & 25.29 & 3.07 & 2.30 & 58.62 & 1.53 \\
DualAvgCode & 25.63 & 22.61 & 3.52 & 28.64 & 19.10 & 0.50 \\
ParityShift & 7.02 & 29.39 & 6.58 & 3.95 & 52.19 & 0.88 \\
WordShift & 29.17 & 22.92 & 2.08 & 25.42 & 20.00 & 0.42 \\
\bottomrule
\end{tabular}}
\end{table*}

\begin{table*}[h!]
\centering
\caption{Error Type Percentages for Different Algorithms in DeepSeek-V3 Model}
\label{tab:error_3}
\resizebox{\textwidth}{!}{
\begin{tabular}{c|cccccc}
\toprule
\multirow{2}{*}{Algorithm} & \multicolumn{6}{c}{Error Types} \\
& Omission/Insertion & Name Decryption Error & Semantic Inference & Reorganization & Reasoning Failure & Other \\
\midrule
Rot13 & 10.73 & 55.93 & 15.82 & 5.08 & 11.86 & 0.56 \\
Atbash & 8.07 & 38.12 & 7.17 & 3.59 & 41.26 & 1.79 \\
Polybius & 5.47 & 12.11 & 2.34 & 2.73 & 76.95 & 0.39 \\
Vigenère & 0.38 & 20.77 & 2.69 & 0.77 & 74.23 & 1.15 \\
Reverse & 21.50 & 40.19 & 5.61 & 13.55 & 18.22 & 0.93 \\
SwapPairs & 1.92 & 18.39 & 2.68 & 0.38 & 76.25 & 0.38 \\
DualAvgCode & 3.07 & 12.64 & 3.45 & 2.68 & 77.78 & 0.38 \\
ParityShift & 1.93 & 28.57 & 3.86 & 0.77 & 64.48 & 0.39 \\
WordShift & 27.80 & 29.46 & 4.56 & 17.01 & 20.33 & 0.83 \\
\bottomrule
\end{tabular}}
\end{table*}

\begin{table*}[h!]
\centering
\caption{Error Type Percentages for Different Algorithms in gemini-1.5-pro Model}
\label{tab:error_4}
\resizebox{\textwidth}{!}{
\begin{tabular}{c|cccccc}
\toprule
\multirow{2}{*}{Algorithm} & \multicolumn{6}{c}{Error Types} \\
& Omission/Insertion & Name Decryption Error & Semantic Inference & Reorganization & Reasoning Failure & Other \\
\midrule
Rot13 & 12.98 & 58.02 & 0.76 & 5.34 & 22.14 & 0.76 \\
Atbash & 1.15 & 15.00 & 3.08 & 0.77 & 78.85 & 1.15 \\
Polybius & 4.21 & 17.24 & 3.07 & 1.92 & 71.65 & 1.92 \\
Vigenère & 2.29 & 14.89 & 3.44 & 0.76 & 78.63 & 0.00 \\
Reverse & 20.85 & 33.19 & 8.94 & 10.21 & 26.38 & 0.43 \\
SwapPairs & 6.49 & 25.57 & 1.91 & 1.53 & 63.36 & 1.15 \\
DualAvgCode & 2.68 & 13.03 & 4.60 & 1.92 & 77.39 & 0.38 \\
ParityShift & 3.08 & 28.46 & 3.08 & 0.38 & 64.23 & 0.77 \\
WordShift & 34.25 & 24.20 & 2.74 & 18.72 & 19.63 & 0.46 \\
\bottomrule
\end{tabular}}
\end{table*}

\begin{table*}[h!]
\centering
\caption{Error Type Percentages for Different Algorithms in o1-mini Model}
\label{tab:error_5}
\resizebox{\textwidth}{!}{
\begin{tabular}{c|cccccc}
\toprule
\multirow{2}{*}{Algorithm} & \multicolumn{6}{c}{Error Types} \\
& Omission/Insertion & Name Decryption Error & Semantic Inference & Reorganization & Reasoning Failure & Other \\
\midrule
Rot13 & 26.95 & 38.30 & 13.48 & 17.02 & 1.42 & 2.84 \\
Atbash & 37.35 & 31.33 & 7.23 & 16.87 & 6.02 & 1.20 \\
Polybius & 30.94 & 32.37 & 1.44 & 25.18 & 8.63 & 1.44 \\
Vigenère & 0.00 & 21.43 & 10.71 & 3.57 & 64.29 & 0.00 \\
Reverse & 12.70 & 29.10 & 8.20 & 32.38 & 17.21 & 0.41 \\
SwapPairs & 1.91 & 9.54 & 1.53 & 0.00 & 86.64 & 0.38 \\
DualAvgCode & 0.00 & 18.52 & 0.00 & 3.70 & 77.78 & 0.00 \\
ParityShift & 4.55 & 34.30 & 3.31 & 4.96 & 52.48 & 0.41 \\
WordShift & 11.58 & 28.57 & 4.63 & 5.79 & 49.03 & 0.39 \\
\bottomrule
\end{tabular}}
\end{table*}

\begin{table*}[h!]
\centering
\caption{Error Type Percentages for Different Algorithms in o1 Model}
\label{tab:error_6}
\resizebox{\textwidth}{!}{
\begin{tabular}{c|cccccc}
\toprule
\multirow{2}{*}{Algorithm} & \multicolumn{6}{c}{Error Types} \\
& Omission/Insertion & Name Decryption Error & Semantic Inference & Reorganization & Reasoning Failure & Other \\
\midrule
Rot13 & 16.19 & 28.57 & 4.76 & 5.71 & 43.81 & 0.95 \\
Atbash & 29.09 & 49.09 & 5.45 & 10.91 & 3.64 & 1.82 \\
Polybius & 40.91 & 28.79 & 6.06 & 10.61 & 12.12 & 1.52 \\
Vigenère & 4.62 & 36.15 & 1.54 & 1.15 & 56.15 & 0.38 \\
Reverse & 16.14 & 25.56 & 3.59 & 14.35 & 38.57 & 1.79 \\
SwapPairs & 5.26 & 31.58 & 5.26 & 5.26 & 52.63 & 0.00 \\
DualAvgCode & 24.62 & 33.85 & 3.08 & 2.31 & 35.38 & 0.77 \\
ParityShift & 4.04 & 26.77 & 4.55 & 2.02 & 62.12 & 0.51 \\
WordShift & 30.88 & 24.26 & 2.94 & 18.38 & 21.32 & 2.21 \\
\bottomrule
\end{tabular}}
\end{table*}

\textbf{Tables~\ref{tab:error_1}--\ref{tab:error_6}} present a detailed breakdown of error distributions across different encryption algorithms for the six selected models. From these results, we identify several common trends and model-specific differences.

\noindent{\textbf{Challenges in Name Decryption and Symbolic Reasoning.}}  
Across all models, name decryption errors remain prevalent, particularly in Atbash and Polybius, indicating persistent difficulties in handling entity-based transformations. Additionally, models struggle with key-based and transposition ciphers such as Vigenère and SwapPairs, suggesting limitations in tracking multi-step transformations and generalizing decryption strategies.

\noindent{\textbf{Semantic Overreliance vs. Overthinking in Decryption.}}  
Chat models often exhibit semantic inference errors, where decrypted outputs align with linguistic patterns rather than encryption rules. In contrast, reasoning models tend to overthink simple tasks, leading to unnecessary self-correction loops that degrade performance in straightforward ciphers like Reverse.

\noindent{\textbf{Structural Alignment and Insertion Errors.}}  
Frequent omission and insertion errors in WordShift and Reverse ciphers highlight difficulties in preserving character order. This suggests that models rely on semantic priors rather than strict symbolic reasoning, leading to misaligned outputs.

\textbf{Key Takeaways:}
\begin{itemize}
    \item Chat models (Claude, Gemini) perform well in substitution ciphers but struggle with complex rule-based encryption.
    \item Reasoning models (DeepSeek-R1, o1) maintain better structural accuracy but underperform in transposition-based and key-dependent ciphers.
    \item All models show high name decryption errors and reasoning failures in Vigenère and SwapPairs, highlighting gaps in symbolic reasoning and long-term dependency tracking.
\end{itemize}

These observations reveal that no single model excels across all ciphers, emphasizing the need for advancements in structured reasoning and symbolic manipulation for decryption tasks.\textbf{ Future improvements }could focus on:

\begin{itemize}
    \item Minimizing the Impact of Semantic Bias in Logical Inference:  
    Cryptographic reasoning tasks often necessitate abstract rule extraction rather than reliance on semantic interpretation. An excessive dependence on linguistic priors can impede the model’s ability to identify underlying structural transformations, resulting in systematic errors. Future advancements should focus on reducing semantic interference to improve the extraction of abstract logical patterns.

    \item Enhancing Comparative Reasoning for Pattern Recognition:  
    While many decryption tasks in CipherBank are straightforward for humans, models frequently fail to derive correct transformation rules from provided exemplars. Strengthening contrastive reasoning mechanisms can enable models to better differentiate encryption structures, facilitating more effective pattern recognition and decryption.

    \item Addressing Overthinking in Model Reasoning:  
    Experimental results indicate that reasoning models exhibit superior performance on complex tasks but underperform on simpler problems. Analysis of inference trajectories reveals a tendency toward recursive self-evaluation, where models continuously revise their approach, even when a straightforward solution is available. For example, in the Reverse cipher, models occasionally attempt unnecessarily complex reasoning paths instead of applying direct positional transformations. Mitigating such overthinking behaviors could enhance efficiency and robustness in logical reasoning.
\end{itemize}

Addressing these limitations will bridge the gap between linguistic fluency and structured cryptographic reasoning, making LLMs more robust in real-world encryption scenarios.

\end{document}